\pdfoutput=1

\documentclass[11pt,twoside,a4paper,cmspaper,final,collab]{cms-tdr}
\def\svnVersion{466098P}\def\cmsCernNoTag{CERN-EP-2018-038}\def\cmsCernDate{\today}\def\cmsMessage{Published in the Journal of High Energy Physics as \href{http://dx.doi.org/10.1007/JHEP06(2018)101}{\doi{10.1007/JHEP06(2018)101.}}}

\begin{document}\cmsNoteHeader{HIG-17-022}

\hyphenation{had-ron-i-za-tion}
\hyphenation{cal-or-i-me-ter}
\hyphenation{de-vices}
\RCS$Revision: 459230 $
\RCS$HeadURL: svn+ssh://svn.cern.ch/reps/tdr2/papers/HIG-17-022/trunk/HIG-17-022.tex $
\RCS$Id: HIG-17-022.tex 459230 2018-05-09 11:05:33Z dsalerno $

\newcommand{\lumifh}{\ensuremath{35.9\fbinv}\xspace}
\newcommand{\ttH}{\ensuremath{\ttbar \PH}\xspace}
\newcommand{\Hbb}{\ensuremath{\PH \to \bbbar}\xspace}
\newcommand{\tthf}{\ensuremath{\ttbar \text{+hf}}\xspace}
\newcommand{\ttlf}{\ensuremath{\ttbar \text{+lf}}\xspace}
\newcommand{\ttjets}{\ensuremath{\ttbar \text{+jets}}\xspace}
\newcommand{\ttbb}{\ensuremath{\ttbar \text{+} \bbbar}\xspace}
\newcommand{\ttb}{\ensuremath{\ttbar \text{+} \cPqb}\xspace}
\newcommand{\tttwob}{\ensuremath{\ttbar \text{+} 2\cPqb}\xspace}
\newcommand{\ttcc}{\ensuremath{\ttbar \text{+} \ccbar}\xspace}
\newcommand{\ttheavy}{\ensuremath{\ttbar \text{+heavy-flavour}}\xspace}
\newcommand{\ttlight}{\ensuremath{\ttbar \text{+light-flavour}}\xspace}
\newcommand{\ttV}{\ensuremath{\ttbar \text{+V}}\xspace}
\newcommand{\ttW}{\ensuremath{\ttbar \text{+} \PW}\xspace}
\newcommand{\ttZ}{\ensuremath{\ttbar \text{+} \cPZ}\xspace}
\newcommand{\Vjets}{\ensuremath{\text{V+jets}}\xspace}
\newcommand{\Wjets}{\ensuremath{\PW \text{+jets}}\xspace}
\newcommand{\Zjets}{\ensuremath{\cPZ \text{+jets}}\xspace}
\newcommand{\massHiggs}{\ensuremath{m_{\PH}}\xspace}
\newcommand{\masstop}{\ensuremath{m_{\cPqt}}\xspace}
\newcommand{\mqq}{\ensuremath{m_{\cPq \cPq}}\xspace}
\newcommand{\muhat}{\ensuremath{\hat{\mu}}\xspace}
\newcommand{\tyukawa}{\ensuremath{y_\cPqt}\xspace}

\newlength\cmsFigWidth
\setlength\cmsFigWidth{0.4\textwidth}
\newlength\cmsTabSkip
\setlength\cmsTabSkip{1.5ex}

\cmsNoteHeader{HIG-17-022}
\title{Search for \ttH production in the all-jet final state in proton-proton collisions at $\sqrt{s}=13\TeV$}

\date{\today}

\abstract{
 	  A search is presented for the associated production of a Higgs boson with a top quark pair in the all-jet final state. Events containing seven or more jets are selected from a sample of proton-proton collisions at $\sqrt{s} = 13\TeV$ collected with the CMS detector at the LHC in 2016, corresponding to an integrated luminosity of \lumifh. To separate the \ttH signal from the irreducible \ttbb background, the analysis assigns leading order matrix element signal and background probability densities to each event. A likelihood-ratio statistic based on these probability densities is used to extract the signal. The results are provided in terms of an observed \ttH signal strength relative to the standard model production cross section $\mu=\sigma/\sigma_\mathrm{SM}$, assuming a Higgs boson mass of 125\GeV.
	  The best fit value is $\muhat = 0.9 \pm 0.7\stat \pm 1.3\syst = 0.9 \pm 1.5 \,(\text{tot})$, and the observed and expected upper limits are, respectively, $\mu < 3.8$ and $< 3.1$ at 95\% confidence levels.
}

\hypersetup{%
pdfauthor={CMS Collaboration},%
pdftitle={Search for ttH production in the all-jet final state in proton-proton collisions at sqrt(s) = 13 TeV},%
pdfsubject={CMS},%
pdfkeywords={CMS, physics, Higgs production, ttbar}}

\maketitle 

\section{Introduction}
\label{sec:intro}
Since the observation of a new boson with a mass of approximately
125\GeV~\cite{ATLAS:2012gk,CMS:2012gu,Chatrchyan:2013lba}, a major focus of the
program of the ATLAS and CMS experiments at the CERN LHC has been to measure the properties
of this particle. To date, all measurements remain consistent with the standard model (SM) Higgs boson (\PH)
hypothesis. The new particle has been observed in
$\gamma\gamma$, $\cPZ\cPZ$, $\PW\PW$, and $\tau\tau$ decays~\cite{Aad:2014eha, Khachatryan:2014ira, Aad:2014eva, Chatrchyan:2013mxa, Aad:2015ona, Chatrchyan:2013iaa, Sirunyan:2017khh}, and evidence has been reported in \bbbar
final states~\cite{Aaboud:2017xsd, Sirunyan:2017elk}.
The observed production and decay rates and spin-parity properties are compatible
with SM expectations~\cite{Aad:2015gba, Khachatryan:2014jba, Chatrchyan:2012jja, Aad:2013xqa, Khachatryan:2016vau}
and with the measured mass of $m_{\PH} = 125.26 \pm 0.21\GeV$~\cite{Sirunyan:2017exp}.

This paper describes a search for the SM Higgs boson produced in
association with a pair of top quarks (\ttH production) in all-jet events,
where the Higgs boson decays exclusively to \bbbar, and each top quark decays
to a bottom quark and a {\PW} boson, which in turn decays to two quarks.
The analysis uses $\sqrt{s}=13\TeV$ proton-proton ({\Pp\Pp}) collision data
collected by the CMS detector in 2016, corresponding to an integrated luminosity of \lumifh.

In the context of probing the properties of the Higgs boson, the \ttH process is
interesting for several reasons. Because of the large top quark mass of $\masstop=172.44 \pm 0.48\GeV$~\cite{Khachatryan:2015hba},
the top quark Yukawa coupling (\tyukawa) is close to unity, and it therefore plays a major role in
many beyond the SM theories. Although model-dependent indirect
measurements of \tyukawa have been made in processes involving top quark loops,
such as in Higgs boson production
through gluon-gluon fusion and in Higgs boson decays to photons, these loop
processes can be influenced by new contributions.
A direct measurement of \tyukawa is therefore crucial to constrain many beyond the SM models.
As the direct measurements of \tyukawa have not yet reached suitable precision in specific final-state
searches nor in their combinations, the addition of an independent search will enhance
the overall sensitivity of such measurements.

The ATLAS and CMS experiments have previously searched for
\ttH production in data corresponding to $\approx$5, 20, and
$36\fbinv$ at centre-of-mass energies of 7, 8, and $13\TeV$,
respectively~\cite{Aad:2014eva, Aaboud:2017jvq, Khachatryan:2014qaa, Sirunyan:2018shy}.
The most recent searches in the \Hbb decay mode
have been performed in the leptonic channels by CMS~\cite{Khachatryan:2015ila} and ATLAS~\cite{Aaboud:2017rss}.
In addition, ATLAS dedicated a search for \ttH in the
all-jet final state at $\sqrt{s}=8\TeV$, in which the observed and expected upper limits were set at
multiples of 6.4 and 5.4 of the respective SM cross sections ($\sigma_\text{SM}$) at 95\% confidence levels (\CL),
and a best fit value was obtained for the signal strength relative to the SM value of $\muhat = \sigma/\sigma_\text{SM} = 1.6 \pm 2.6$~\cite{Aad:2016zqi}.

\begin{figure}[hbtp]
  \centering
   \includegraphics[width=0.9\cmsFigWidth]{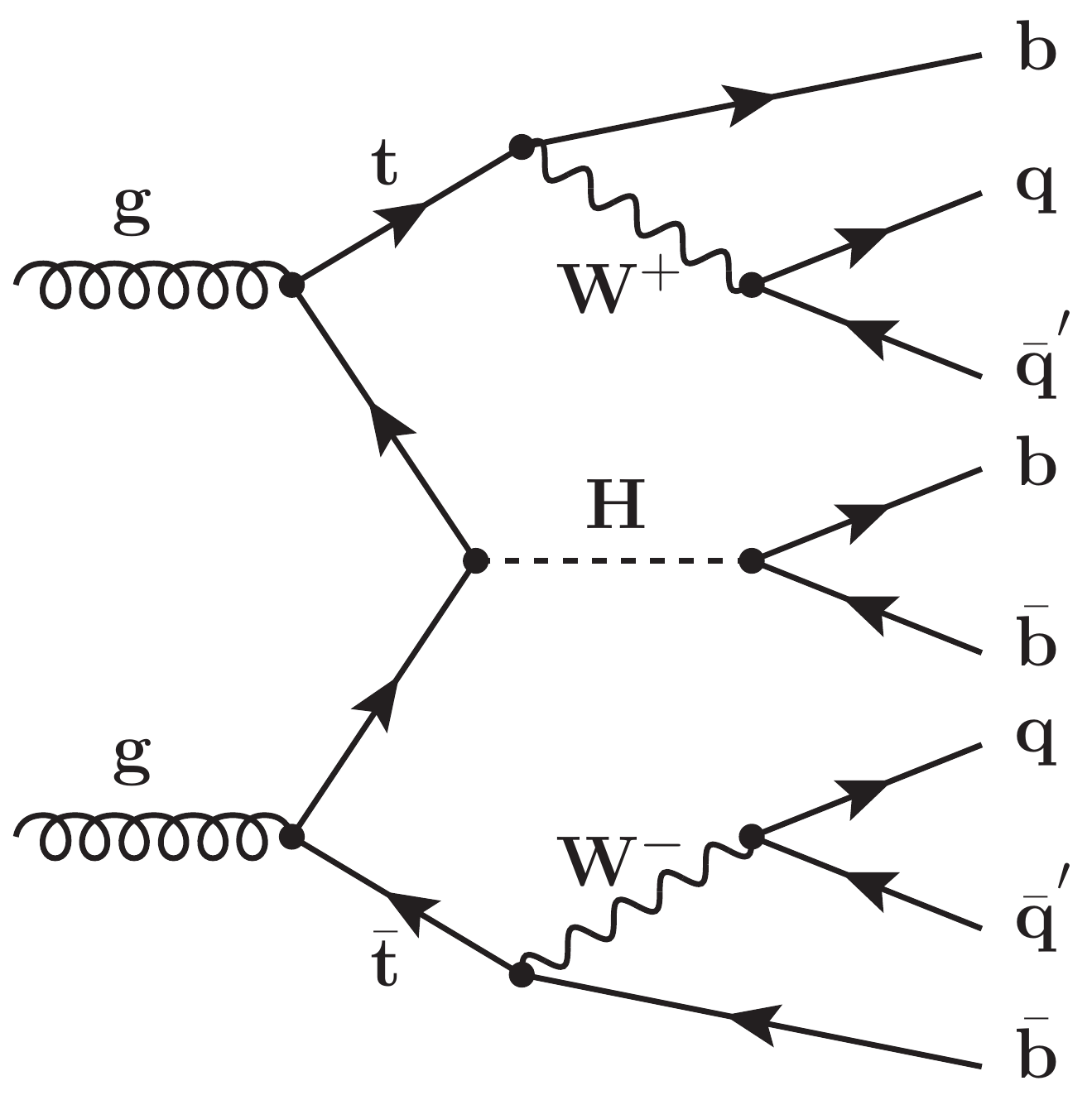}
   \caption{An example of an LO Feynman diagram for \ttH
     production, including the subsequent decay of the top
     quark-antiquark pair, as well as that of the Higgs boson
     into a bottom quark-antiquark pair.}
   \label{fig:ttHbb_AH_feynman}
\end{figure}

The leading order (LO) Feynman diagram for \ttH production in the all-jet final state
is shown in Fig.~\ref{fig:ttHbb_AH_feynman}.
Ideally, signal events contain eight jets,
of which four are tagged as originating from {\cPqb} quarks via the algorithm described
in Section~\ref{sec:selection}. To accommodate jets lost due to
geometrical acceptance, merging of jets, and
{\cPqb} tagging inefficiency, events with seven or more jets and three or more
{\cPqb}-tagged jets are included in the defined signal region (SR). To account for extra
jets from initial- or final-state radiation, up to nine jets are considered per event.

Despite that the signature involves a large number of
high transverse momentum (\pt) final-state jets,
the analysis is constrained by background, mainly from jets produced through the strong interaction,
referred to as quantum chromodynamic (QCD) multijet production.
The next largest background contribution is from \ttjets,
which includes \ttlight jets, where one or more of the jets are incorrectly
identified as {\cPqb} jets, \ttcc, and the topologically irreducible \ttbb
background. Smaller background contributions arise from single top quark
and vector boson production in association with jets.
Although the all-jet final state has significantly more background than, for example the lepton+jets final state,
it offers a fully reconstructed event, and a larger contribution from the signal.

Because background rates are much larger than the
\ttH signal, it is important that all available information on the differences between
the signal and backgrounds is incorporated into the analysis.
A quark-gluon discriminant is used to differentiate between events
containing jets originating from light-flavour quarks and events
containing jets from gluons.
Given the many combinations of jet--quark associations, it is difficult
to resolve a clear, resonant Higgs boson mass peak. Nevertheless,
there are underlying kinematic differences between the \ttH signal
and the multijet background and, to a lesser extent, the \ttjets background.
These differences are exploited through
the matrix element method (MEM)~\cite{Abazov:2004cs, Abazov:2004ym, Aaltonen:2009dh, Khachatryan:2015ila}
to distinguish the signal from background.
Specifically, events are assigned a
probability density according to how compatible they are with the \ttH process.
Although this probability density suffices to separate the signal from most backgrounds,
a second probability density is assigned to each event according to its
compatibility with the \ttbb process, which provides additional discrimination
against the topologically irreducible \ttbb background. The two probability densities are combined
in a likelihood ratio to form the final discriminant between signal and background.

The body of this paper is devoted to the details of the analysis. The
CMS detector is described in Section~\ref{sec:detector}, while the
description of the Monte Carlo (MC) simulation follows in
Section~\ref{sec:samples}. The reconstruction of physical objects
and event selections are discussed in Section~\ref{sec:selection}.
The QCD multijet background, estimated from control regions in
data, is described in
Section~\ref{sec:qcd}, followed by the description of the analysis
strategy and the MEM in Section~\ref{sec:strategy}.
The treatment of the systematic uncertainties is discussed in Section~\ref{sec:syst},
the results are presented in Section~\ref{sec:result}, and a brief summary is
provided in Section~\ref{sec:summary}.

\section{The CMS detector}
\label{sec:detector}
The central feature of the CMS apparatus is a superconducting solenoid of 6\unit{m} internal diameter,
providing a magnetic field of 3.8\unit{T}. A
silicon pixel and strip tracker, a lead tungstate crystal electromagnetic calorimeter (ECAL), and
a brass and scintillator hadron calorimeter (HCAL), each composed of a barrel and two end
sections, reside within the field volume. Forward calorimeters extend the pseudorapidity ($\eta$) coverage beyond these barrel and
end sections. Muons are measured in gas-ionization detectors embedded in the steel flux-return
yoke outside the solenoid.
Events of interest are selected using a two-tiered trigger system~\cite{Khachatryan:2016bia}.
The first level, composed of specialized hardware processors, uses information from the calorimeters
and muon detectors to select events at a rate close to 100\unit{kHz}, ascertained within a time interval of
less than 4\mus. The second level, known as the high-level trigger, consists of a farm of
processors running a version of the full event reconstruction software optimized for fast processing,
and reduces the event rate to $\approx$1\unit{kHz} before data storage.
A more detailed description of the CMS detector, together
with a definition of the coordinate system and the kinematic variables, can be
found in Ref.~\cite{Chatrchyan:2008aa}.

\section{Simulated samples}
\label{sec:samples}
This analysis uses MC event generators and a detailed detector simulation based on \GEANTfour (v9.4)~\cite{Agostinelli:2002hh} to model experimental effects, such as reconstruction, selection efficiencies, and resolutions in the CMS detector.

The \ttH signal is modelled at next-to-LO (NLO) perturbative QCD via the \POWHEG (v2) event generator~\cite{Nason:2004rx, Frixione:2007vw, Alioli:2010xd, Hartanto:2015uka}. The value of the Higgs boson mass is set to $\massHiggs=125\GeV$, the value of the top quark mass to $\masstop=172.5\GeV$, and the parton distribution functions (PDFs) of the proton are described using NNPDF3.0~\cite{Ball:2014uwa}, at the same level of accuracy in perturbative QCD as used to generate the event samples. These events are subsequently processed through \PYTHIA (v8.200)~\cite{Sjostrand:2014zea} for parton showering and hadronization.

The \POWHEG generator~\cite{Campbell:2014kua, Alioli:2009je, Re:2010bp} is used to simulate \ttbar and single top quark (single {\cPqt}) production at NLO. The associated production of \ttbar with a vector boson (referred to as \ttW and \ttZ, or commonly as \ttV in the following) is simulated at NLO using \MGvATNLO (v2.2.2)~\cite{Alwall:2014hca} with FxFx jet matching and merging~\cite{Frederix:2012ps}. The \MGvATNLO generator is also used at LO accuracy with the MLM matching scheme~\cite{Alwall:2007fs} to simulate the production of {\PW} and {\cPZ} bosons with additional jets (\Vjets) as well as QCD multijet events. The multijet background is estimated from control regions in data, as described in Section~\ref{sec:qcd}, while simulated events are used to ensure the self-consistency of the method. Diboson ($\PW\PW$, $\PW\cPZ$, and $\cPZ\cPZ$) events are simulated at LO using \PYTHIA. All background events also use \PYTHIA for parton showering and hadronization.
The underlying event in the simulation of all backgrounds, except for \ttbar, is characterized through the \PYTHIA CUETP8M1 event tune~\cite{Khachatryan:2015pea, Skands:2014pea}.
For \ttH and \ttbar production, we use a custom tune CUETP8M2, developed by CMS with an updated strong coupling $\alpha_s$ for initial-state radiation, to better model the jet multiplicity spectrum.

For comparing with measured distributions, the events in the simulated samples are normalized to the integrated luminosity of the data according to their predicted cross sections, taken from calculations at next-to-NLO (NNLO) for \Vjets production, at approximate NNLO for single {\cPqt} production~\cite{Kidonakis:2013zqa}, and at NLO for diboson~\cite{Campbell:2011bn} and \ttV production~\cite{Maltoni:2015ena}. The \ttH cross section and Higgs boson branching fractions are also generated at NLO accuracy~\cite{deFlorian:2016spz}.
The simulated \ttbar events are normalized to the full NNLO calculation, including soft-gluon resummation at next-to-next-to-leading logarithmic accuracy~\cite{Czakon:2011xx}, assuming $\masstop = 172.5\GeV$. The dependence on choice of the PDFs was also studied and found to have negligible impact on all observed distributions; however, their impact on systematic uncertainties enters through their normalization.
As the differential measurements of top quark pair production indicate that the measured \pt spectrum of top quarks is softer than in simulation~\cite{Khachatryan:2016oou}, the top quark \pt distribution in \POWHEG-generated \ttbar events is reweighted to the data using a control region enriched in lepton+jets \ttbar events.
The events in the \ttbar sample are also separated according to the flavour of additional jets that do not originate from top quark decays, namely: \ttbb, where two additional {\cPqb} jets are within the acceptance defined in Section~\ref{sec:selection} and originate from one or more {\cPqb} quarks; \ttb, where only one {\cPqb} jet is within the acceptance and originates from a single {\cPqb} quark; \tttwob, where a single {\cPqb} jet is within the acceptance and originates from two
spatially close {\cPqb} quarks; \ttcc, where there is at least one {\cPqc} jet and no additional {\cPqb} jets within the acceptance; and \ttlight partons (\ttlf). The separation is motivated by the fact that the subsamples originate from different processes and therefore have different associated systematic uncertainties.

Effects from additional {\Pp\Pp} interactions in the same or neighbouring bunch crossings (referred to as pileup) are modelled by adding simulated minimum-bias events generated in \PYTHIA to all simulated processes. The pileup multiplicity distribution in simulation is reweighted to reflect the one observed in {\Pp\Pp} collision data, which is distributed between about 10 and 40 and peaks at $\approx$20. Additional rescaling factors are applied where necessary to improve the description of data in simulation, as discussed in the next section.

\section{Event reconstruction and selection}
\label{sec:selection}
The particle-flow event algorithm~\cite{Sirunyan:2017ulk} is used to reconstruct and identify physical objects from an optimized combination of information from the various components of the CMS detector.
The reconstructed vertex with the largest value of summed object $\pt^2$ is taken to be the primary {\Pp\Pp} interaction vertex. These objects correspond to jets, clustered with the jet finding algorithm~\cite{Cacciari:2008gp,Cacciari:2011ma} using the tracks assigned to the vertex as inputs, and to the associated missing transverse momentum (\ptvecmiss), defined as the negative vector sum of the \pt of those jets.
The energy of charged hadrons is determined from a combination of their momenta measured in the tracker and the corresponding ECAL and HCAL energy depositions, corrected
for the response function of the calorimeters to hadronic showers, while the energy of neutral hadrons is obtained from the corresponding corrected ECAL and HCAL energies. Charged hadrons from pileup events are omitted in the subsequent event reconstruction. The energy of electrons is determined from a combination of the electron momentum at the primary interaction vertex obtained from the tracker, the energy in the corresponding ECAL clusters, and the energy sum of all bremsstrahlung photons spatially compatible with the electron track. The \pt of muons is based on the curvature of the corresponding track.

The selection criteria reflect the all-jet final state in \ttH production where the Higgs boson decays into a \bbbar pair. The analysis requires the reconstruction of all objects originating from a common vertex, \ie jets from quark hadronization, including {\cPqb}-tagged jets, and electrons and muons to veto any events with leptons. The lepton veto ensures that leptonic final states in \ttH production are not considered, as these are covered by separate searches at CMS.

Jets are reconstructed from particle-flow candidates using the anti-\kt clustering algorithm~\cite{Cacciari:2008gp}, implemented in the \FASTJET~\cite{Cacciari:2011ma} package, using a distance parameter of 0.4. In the offline reconstruction, each jet is required to have $\pt>30\GeV$ and $\abs{\eta}<2.4$, and ECAL and HCAL energy fractions of at least 1\%, which removes jets originating from instrumental effects.
In addition, jet energy corrections are applied that depend on \pt, $\eta$, pileup, and residual differences between data and MC simulations~\cite{Chatrchyan:2011ds, Khachatryan:2016kdb}.

Jets originating from decays of {\cPqb} quarks are identified using the combined secondary vertex (CSVv2) algorithm~\cite{Chatrchyan:2012jua, CMS-BTV-16-002}, which uses
information from the tracker. The algorithm combines information on the impact parameters at the primary vertex with features of secondary vertices within the jet, via a neural network discriminant. The least restrictive (``loose'') and ``medium'' working points of the CSVv2 algorithm have respective efficiencies of $\approx$80 and 65\% for tagging jets originating from {\cPqb} quarks, $\approx$35 and 10\% for jets originating from {\cPqc} quarks, and probabilities of $\approx$10 and 1\% for jets from light-flavour quarks or gluons to be misidentified as {\cPqb} jets. In the remainder of this paper, jets passing the loose and medium {\cPqb} tagging working points are referred to as CSVL and CSVM jets, respectively. The distribution in the output of the CSVv2 discriminant in simulation is corrected using scale factors to provide a better description of the observed features in data. This correction is obtained separately for light-flavour and {\cPqb} jets via a ``tag-and-probe'' approach~\cite{Khachatryan:2010xn} in dilepton control samples enriched in events with a {\cPZ} boson and two jets, and in \ttbar events with no additional jets, respectively. No correction is applied for {\cPqc} jets, although a larger systematic uncertainty is assigned to such jets, as described in Section~\ref{sec:syst}.

Two dedicated triggers have been developed to efficiently select signal events, both requiring at least six jets with $\abs{\eta}<2.6$.
The first trigger has kinematic requirements of jet $\pt>40\GeV$ and $\HT>450\GeV$, and requires at least one of the jets to be {\cPqb} tagged; the second trigger is complementary in that it has less stringent kinematic conditions of jet $\pt>30\GeV$ and $\HT>400\GeV$, but requires at least two {\cPqb}-tagged jets; the \HT being defined by the scalar sum of jet \pt of all jets in the event satisfying the given thresholds.
The {\cPqb} jets are tagged online by the triggers at an efficiency of $\approx$70--80\%, with a misidentification rate of $\approx$6\% for light-flavour quark and gluon jets. Events in data and simulation are selected if they pass either of these two triggers.
The efficiency in data and in simulation is measured in bins of the number of CSVM jets, the \pt of the jet with the sixth-highest \pt, and the \HT in control samples collected using single-muon triggers. A bin-by-bin scale factor, based on the ratio of this efficiency in data to that in simulation, is applied to simulated events to correct for any remaining differences. The overall trigger efficiency for signal events that pass the offline event selection is 99\%.

\begin{figure}[hbtp]
  \centering
   \includegraphics[width=0.49\textwidth]{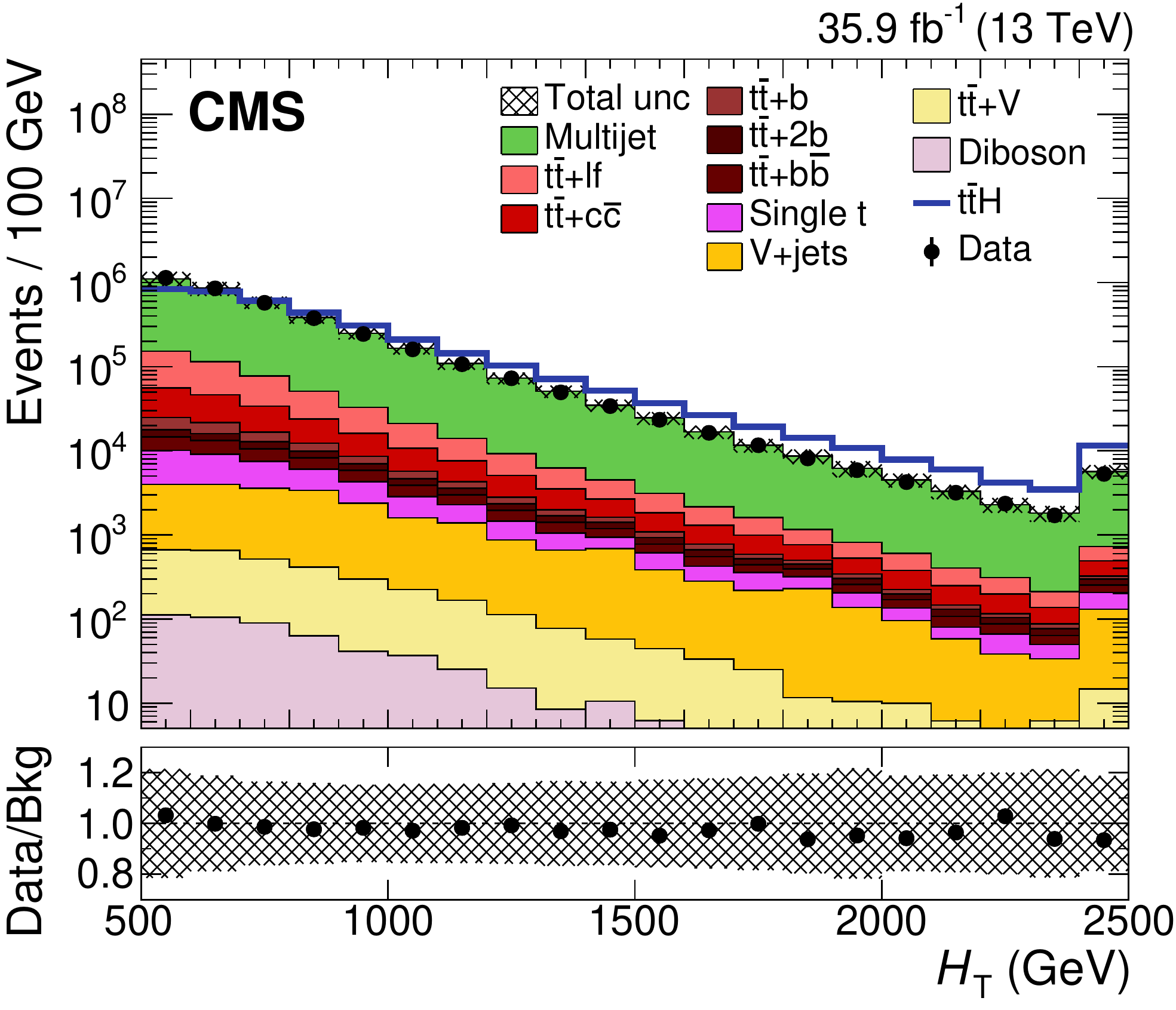} 
      \includegraphics[width=0.49\textwidth]{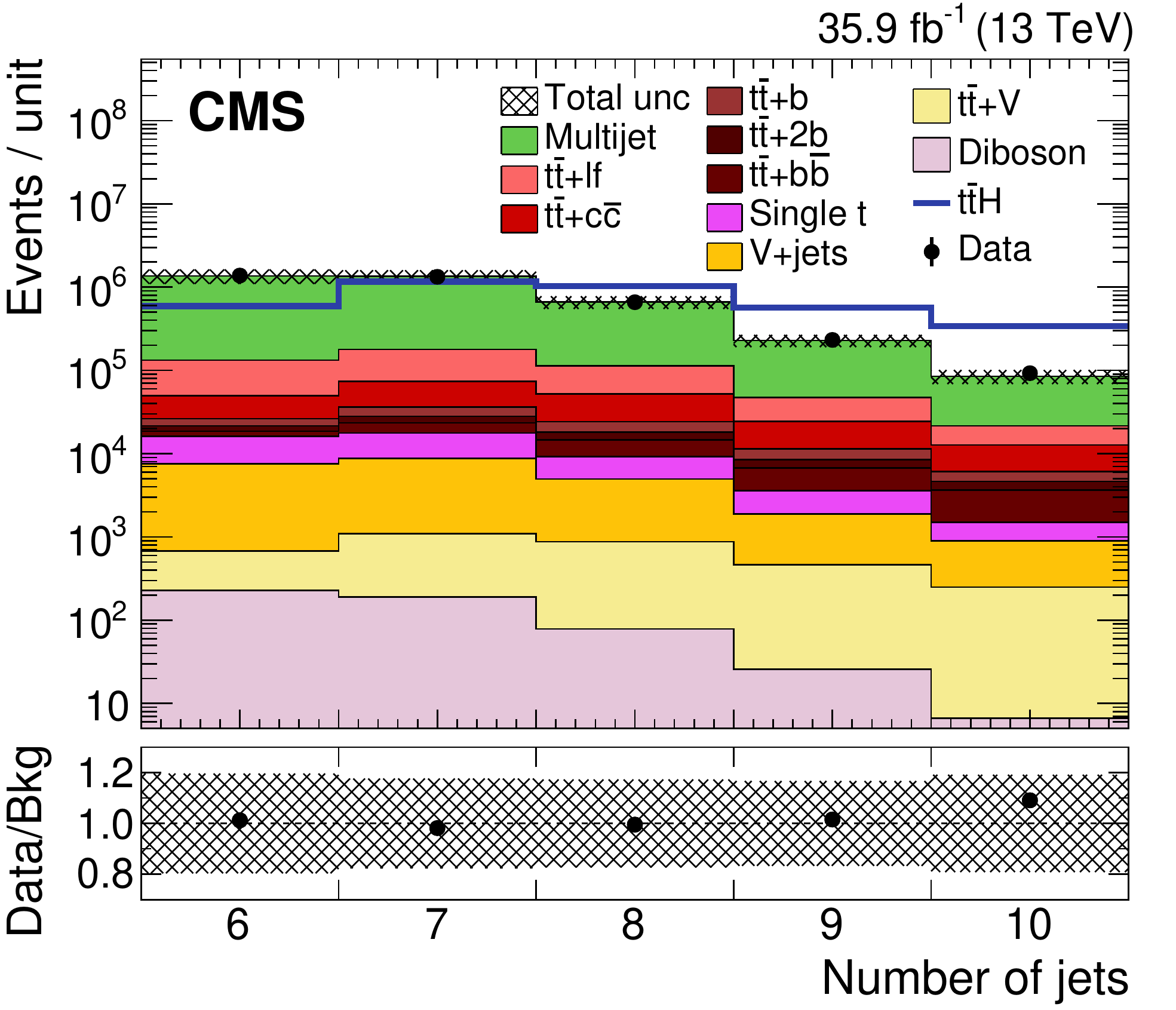}
    \caption{Distribution in \HT (left) and jet multiplicity (right) in data (black points) and in simulation (stacked histograms), after implementing the preselection.
      The simulated backgrounds are first scaled to the integrated luminosity of the data, and then the simulated QCD multijet background is rescaled to match the yield in data.
      The contribution from \ttH signal (blue line) is scaled to the total background yield (equivalent to the yield in data) to enhance readability.
      The hatched bands reflect the total statistical and systematic uncertainties in the background prediction, prior to the fit to data, which are dominated by the systematic uncertainties in the simulated multijet background.
      The last bin includes event overflows.
      The ratios of data to background are given below the main panels.
    }
    \label{fig:objects:ht_nj}
\end{figure}

To ensure that the trigger selection is close to full efficiency relative to the offline selection, thereby reducing the uncertainty in any efficiency differences between data and simulation, the offline analysis selects events that contain at least six jets with $\pt>40\GeV$, requiring $\HT>500\GeV$ and at least 2 {\cPqb}-tagged CSVM jets. Events with reconstructed electrons or muons in the final state are vetoed using loose lepton reconstruction and identification criteria, in particular, electrons and muons are required to have $\pt>15\GeV$, $\abs{\eta}<2.4$, and to be isolated from jet activity.
If at least one reconstructed electron or muon is found in an event, the event is rejected. The distributions in \HT and jet multiplicity after the preselection are shown in Fig.~\ref{fig:objects:ht_nj} for the data and simulation, with the signal contribution scaled to the total background yield for illustration.

After this preselection, the events are categorized according to their reconstructed jet and {\cPqb} jet multiplicities.
Six categories are formed: 7 reconstructed jets, out of which 3 are CSVM jets (7j,~3{\cPqb}); 7 jets with 4 or more CSVM jets (7j,~$\geq$4{\cPqb}); 8 jets with 3 CSVM jets (8j,~3{\cPqb}); 8 jets with 4 or more CSVM jets (8j,~$\geq$4{\cPqb}); 9 or more jets with 3 CSVM jets ($\geq$9j,~3{\cPqb}); and 9 or more jets with 4 or more CSVM jets ($\geq$9j,~$\geq$4{\cPqb}).
Events with 7 or more jets and 2 CSVM jets are used to form control regions for the multijet background estimation, as described in Section~\ref{sec:qcd}.

To reject events that are unlikely to include a {\PW} boson from top quark decay, a cutoff is placed on the dijet invariant mass. Any jets that are not {\cPqb} tagged are considered in this calculation, and the pair with invariant mass \mqq closest to the {\PW} boson mass is chosen as the {\PW} boson candidate in the event. Requirements of $60 < \mqq < 100$, and $70 < \mqq < 92\GeV$ are applied in the 7- or 8-jet, and 9-or-more-jet categories, respectively.

\begin{figure}[hbtp]
  \centering
      \includegraphics[width=0.49\textwidth]{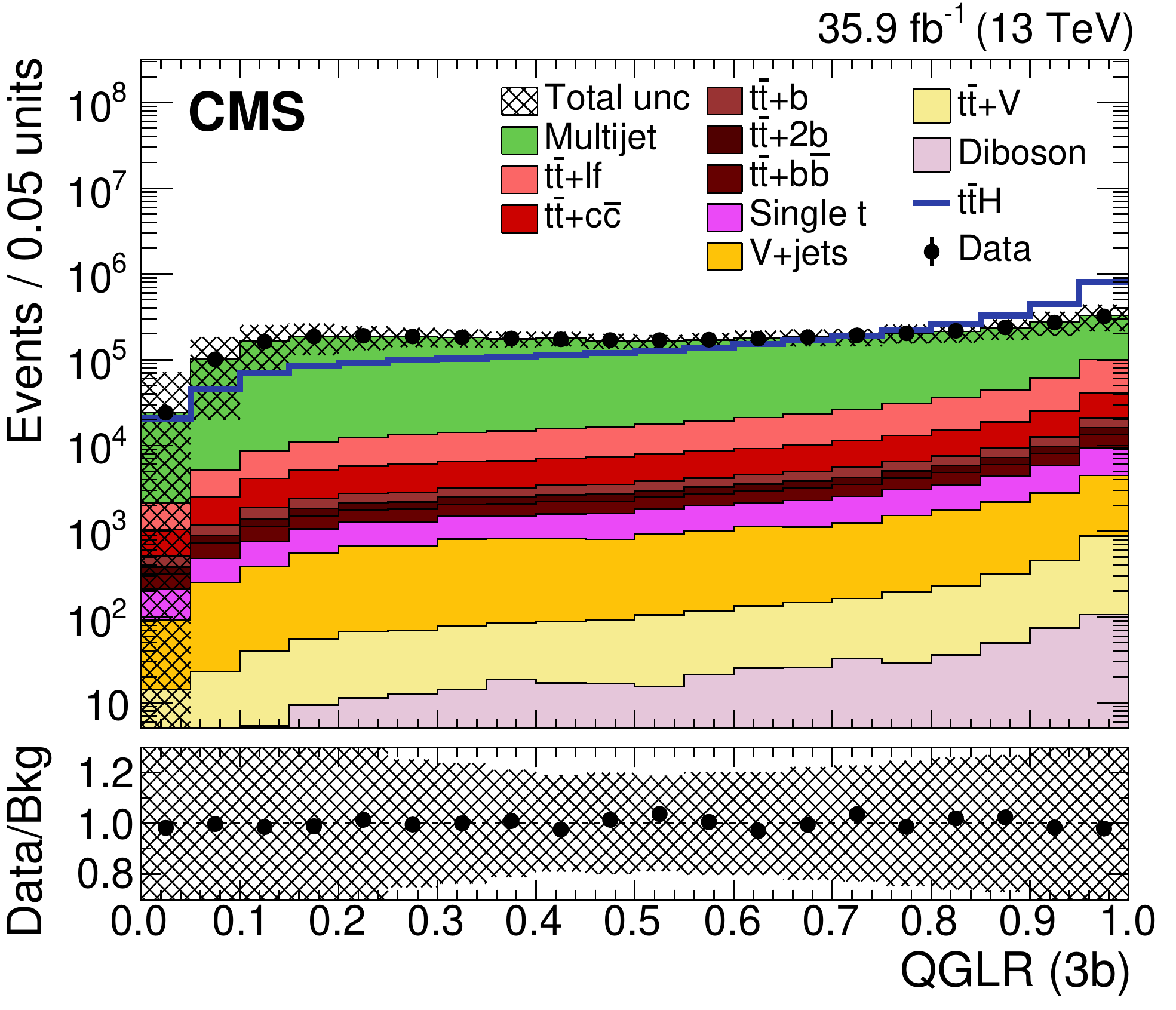}
    \caption{Comparison of the distributions in the quark-gluon likelihood ratio in data (black points) and in simulation (stacked histograms), after preselection, excluding the first 3 {\cPqb}-tagged jets.
       The simulated backgrounds are first scaled to the integrated luminosity of the data, and then the simulated multijet background is rescaled to match the yield in data.
      The contribution from signal (blue line) is scaled to the total background yield (equivalent to the yield in data) to enhance readability.
      The hatched bands reflect the total statistical and systematic uncertainties in the background prediction, prior to the fit to data, which are dominated by the systematic uncertainties in the simulated multijet background.
      The ratio of data to background is given below the main panel.
    }
    \label{fig:objects:qgWeight}
\end{figure}

Jets are also classified according to a quark-gluon likelihood (QGL)~\cite{CMS-PAS-JME-13-002,CMS-DP-2016-070}. The QGL discriminates light-flavour quark jets from gluon jets, exploiting the higher track multiplicity of gluon jets, the lower \pt of their constituents, as well as their less collimated spatial profile. A jet is assigned a QGL discriminant value near 1 if it is quark-like, while gluon-like jets receive a value near 0. This is used to separate events containing light-flavour jets from $\qqbar^\prime$ decays of {\PW} bosons from events containing gluon jets produced in multijet interactions or via initial-state radiation. The QGL discriminant is used only for jets not identified as originating from {\cPqb} quarks, and is combined in a likelihood ratio (LR) defined as:
\begin{equation} \label{eq:qgLR}
q_{\text{LR}(N_1,N_2)} = \frac{L(N_1,0)}{L(N_1,0) + L(N_2,N_1-N_2)},
\end{equation}
where $N_1$ is the number of jets as well as the number of quarks in the all-quark hypothesis, and $N_2$ is the number of quarks in the quark+gluon hypothesis.
The individual likelihoods are defined by:
\begin{equation} \label{eq:qgLR_s}
L(N_{\cPq},N_{\cPg}) =
\sum_\text{perm} \left(  \prod_{ k = i_1}^{i_{N_{\cPq}}} f_{\cPq}(\zeta_{k})
\prod_{ m = i_{N_{\cPq}+1}}^{i_{N_{\cPq}+N_{\cPg}}} f_{\cPg}(\zeta_{m}) \right),
\end{equation}
where $\zeta_{i}$ is the QGL discriminant for the $i^\text{th}$ jet, and $f_{\cPq}$ and $f_{\cPg}$
are the probability density functions for $\zeta_{i}$ when the $i^\text{th}$ jet originates from a quark or gluon, respectively.
The former include {\cPqu}, {\cPqd}, {\cPqs}, and {\cPqc} quarks. The sum in Eq.~\eqref{eq:qgLR_s} runs over all unique permutations in assigning $N_{\cPq}$ jets to quarks and $N_{\cPg}$ jets to gluons. We use $q_{\text{LR}(N,0)}$ to compare the likelihood of $N$ reconstructed light-flavour jets that arise from $N$ quarks, to the likelihood of $N$ reconstructed light-flavour jets that arise from $N$ gluons.
After excluding either the first 3 or 4 CSVM jets in an event (based on the highest CSVv2 value), up to $N$ light-flavour jets, with $N=3$, 4, or 5 are used to calculate the LR, with a requirement of $q_{\text{LR}(N,0)} > 0.5$ set for the SR.
To correct the modelling of the QGL distribution in simulation, a reweighting based on a control sample of $\mu^+\mu^-$+\,jet and dijet events is applied to each event according to the type (quark or gluon) and QGL value of all jets in the event.
The distribution in the quark gluon LR (QGLR), calculated excluding the first 3 {\cPqb}-tagged jets, is shown in Fig.~\ref{fig:objects:qgWeight} for data and simulation for events passing the preselection. Good agreement is observed between prediction and data.

\section{Background estimation}
\label{sec:qcd}
The main backgrounds stem from multijet and \ttbar production associated with additional gluons, light-flavour, charm, or bottom quarks (\ttjets). The background from \ttjets as well as other minor backgrounds (single {\cPqt}, \Vjets, \ttV, and diboson events) are estimated from MC simulation, while the background from multijet events is obtained from control regions in data, as described below. The approach uses a control region (CR) with low {\cPqb} tag multiplicity to estimate the contribution from multijet events in the SR. The CR is enriched in multijet events, and the remaining contribution from other backgrounds (mainly \ttjets) is subtracted using simulation.

The CR is defined by events with two CSVM jets and at least three CSVL jets. We define a validation region (VR) using events with $\mathrm{QGLR}<0.5$. This definition provides four orthogonal regions, summarized in Table~\ref{table:qcd_regions}, from which we can obtain and check the multijet background estimate. The use of the VR relies on the fact that the QGLR and the number of additional CSVL jets are uncorrelated by construction, since only the non-CSVL jets are used in the calculation of the QGLR, except for the rare case of events with 5 or more CSVL jets.

\begin{table}[hbtp]
  \topcaption{Definition and description of the four orthogonal regions in the analysis.}
  \label{table:qcd_regions}
  \centering
    \begin{tabular}{c c c}
      \hline
	 & $N_\text{CSVM} = 2$ & \multirow{2}{*}{$N_\text{CSVM} \ge 3$} \\
	 & $N_\text{CSVL} \ge 3$ & \\
	\hline
	\multirow{2}{*}{$\textrm{QGLR}>0.5$} & CR & SR  \\
	 & (to extract distribution) & (final analysis) \\[\cmsTabSkip]
	\multirow{2}{*}{$\textrm{QGLR}<0.5$} & Validation CR  & VR  \\
	& (to validate distribution) & (comparison with data) \\
      \hline
    \end{tabular}
\end{table}

The four orthogonal regions are used independently in each of the six analysis categories described in Section~\ref{sec:selection}.
For a given variable, the distribution of the multijet events in the SR of each category is estimated from the data in the CR, after subtracting \ttjets and other minor backgrounds.
Since the kinematic properties of {\cPqb}-tagged and untagged jets differ in the CR and SR because of different heavy-flavour composition, corrections as a function of jet \pt, $\eta$, and the minimum distance between the jet and the first two {\cPqb}-tagged jets ($\Delta R_\text{min}$) are applied to the CSVL jets in the CR.
The correction function is obtained from jets in events passing the preselection, excluding the first two jets, ordered according to their CSVv2 discriminants. This is used to reweight the kinematic distributions of CSVL jets to match those of CSVM jets.
The corrected multijet distribution in the CR is scaled to provide an estimate of the distribution in the SR. Specifically, the multijet yield in each category is left floating in the final fit, as discussed in Section~\ref{sec:result}.

\begin{figure}[hbtp]
  \centering
   \includegraphics[width=\cmsFigWidth]{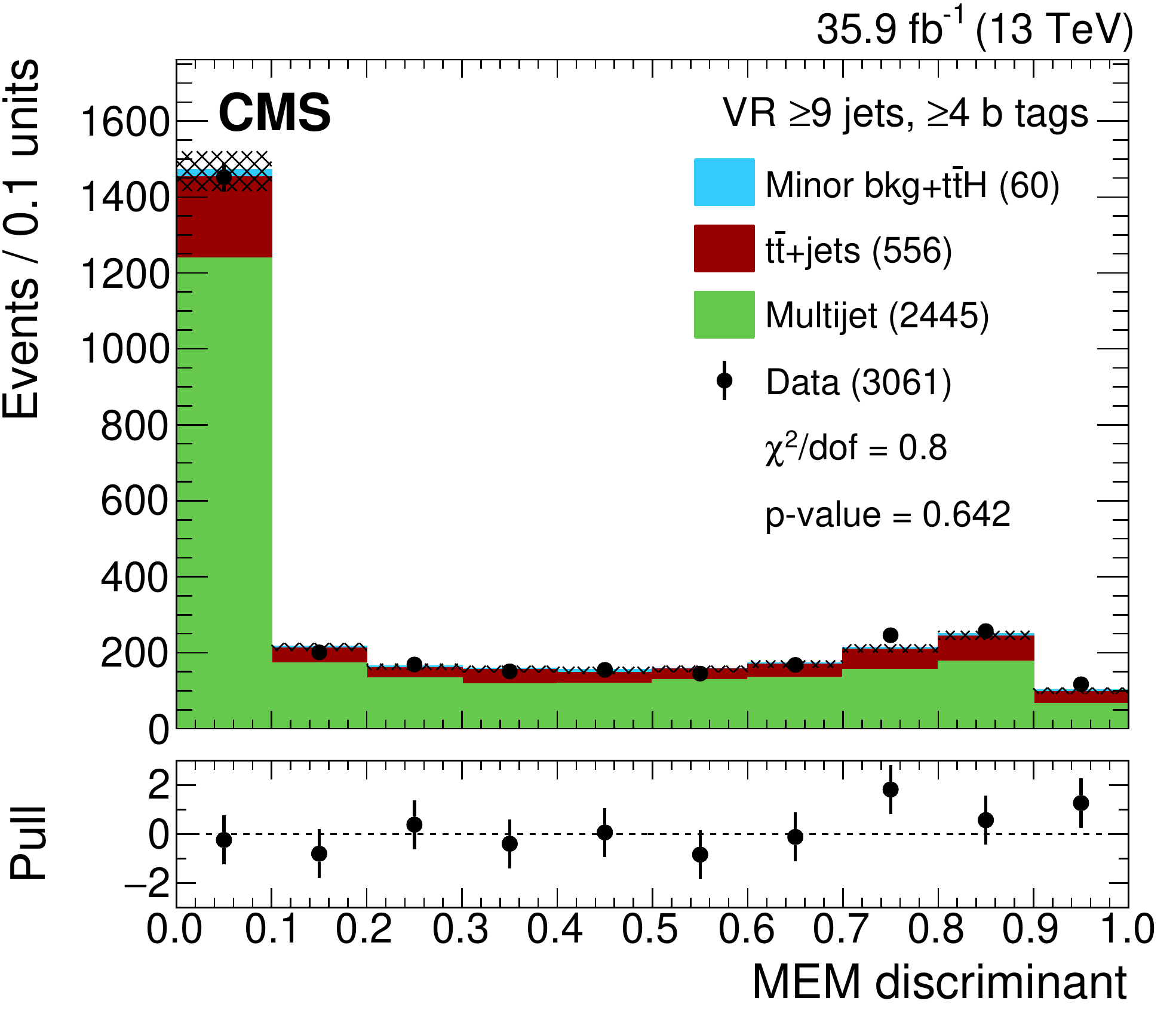}
   \hspace{20pt}
   \includegraphics[width=\cmsFigWidth]{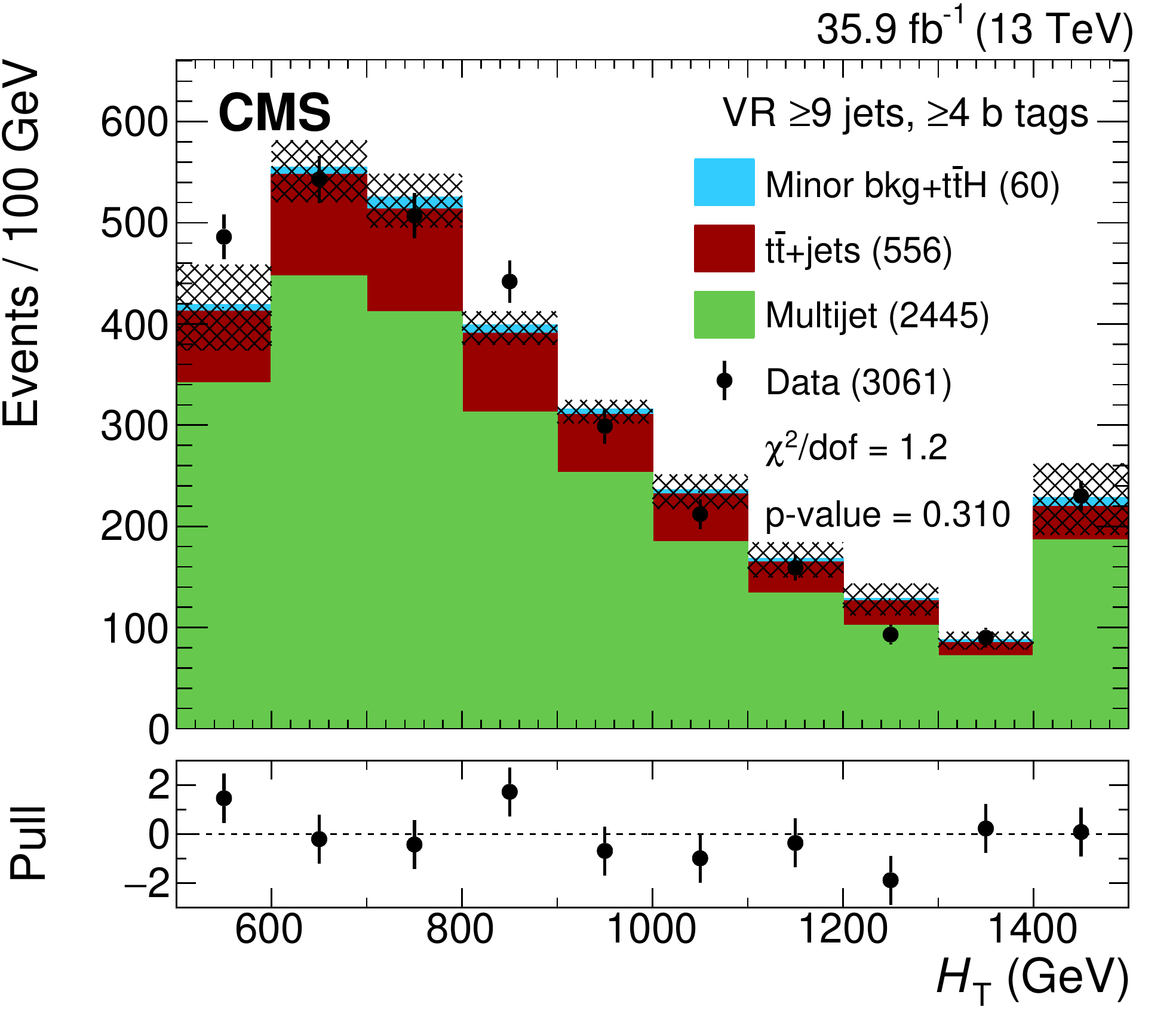} \\
   \vspace{10pt}
   \includegraphics[width=\cmsFigWidth]{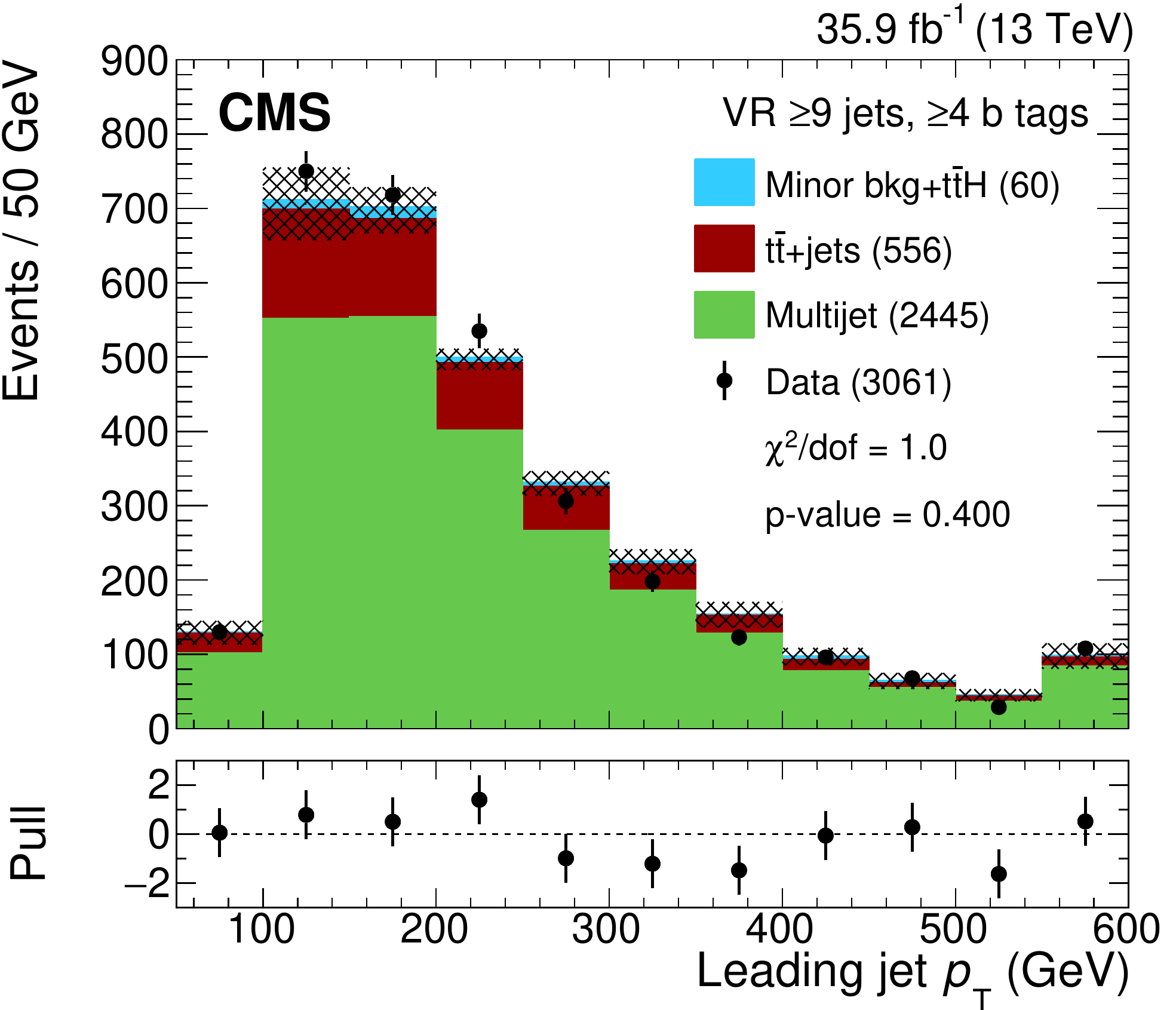}
   \hspace{20pt}
   \includegraphics[width=\cmsFigWidth]{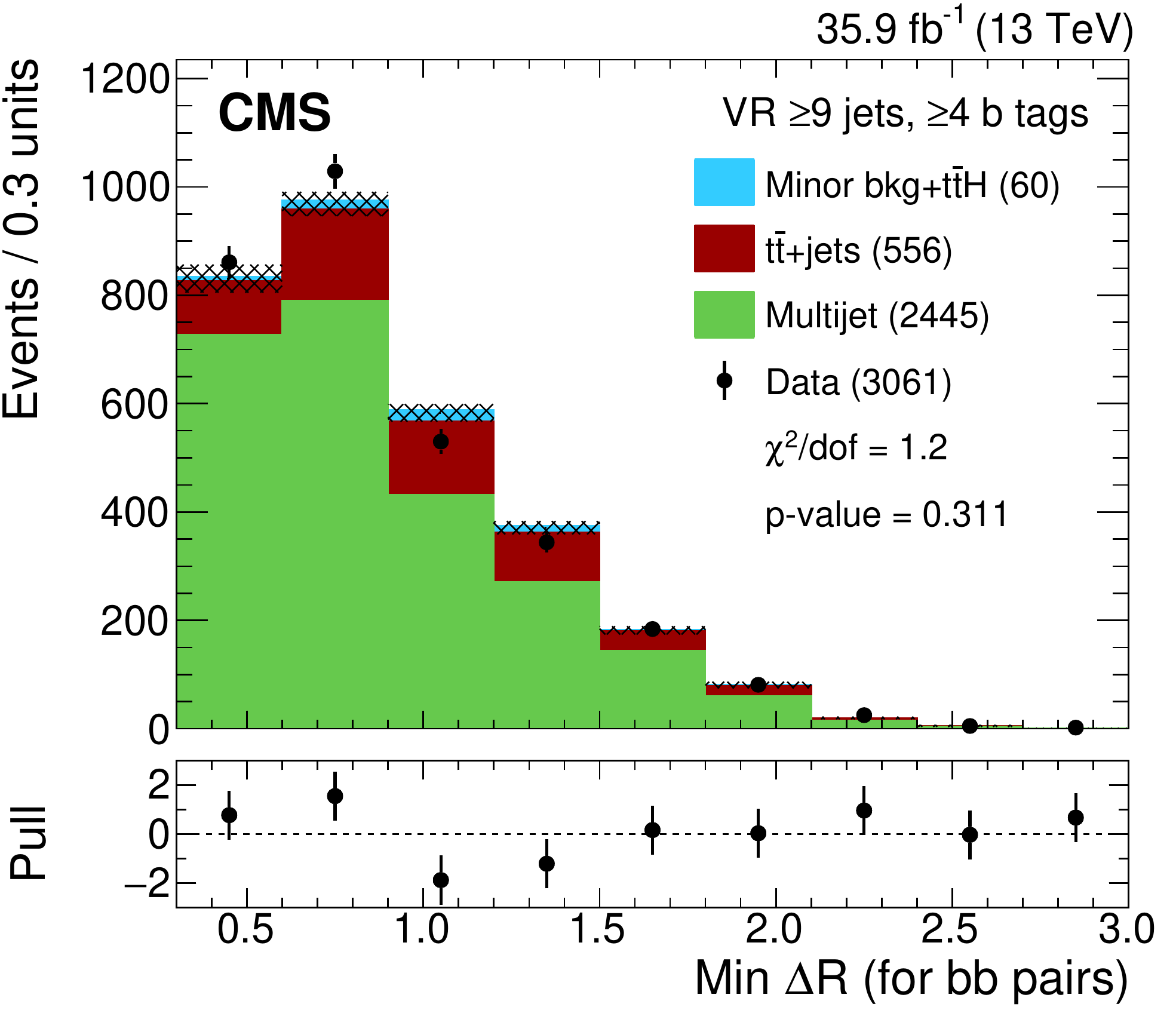} \\
   \vspace{10pt}
   \includegraphics[width=\cmsFigWidth]{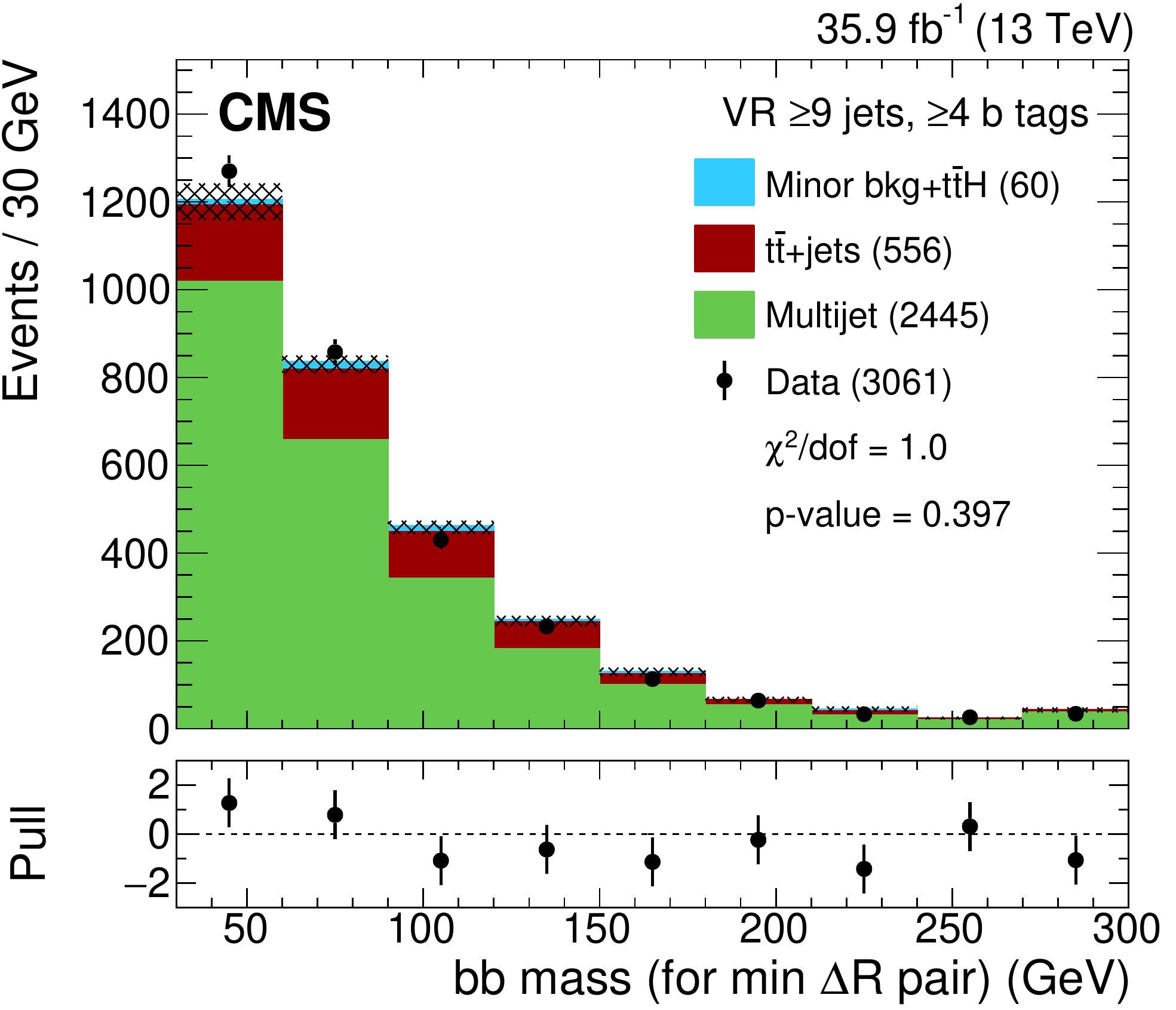}
   \hspace{20pt}
   \includegraphics[width=\cmsFigWidth]{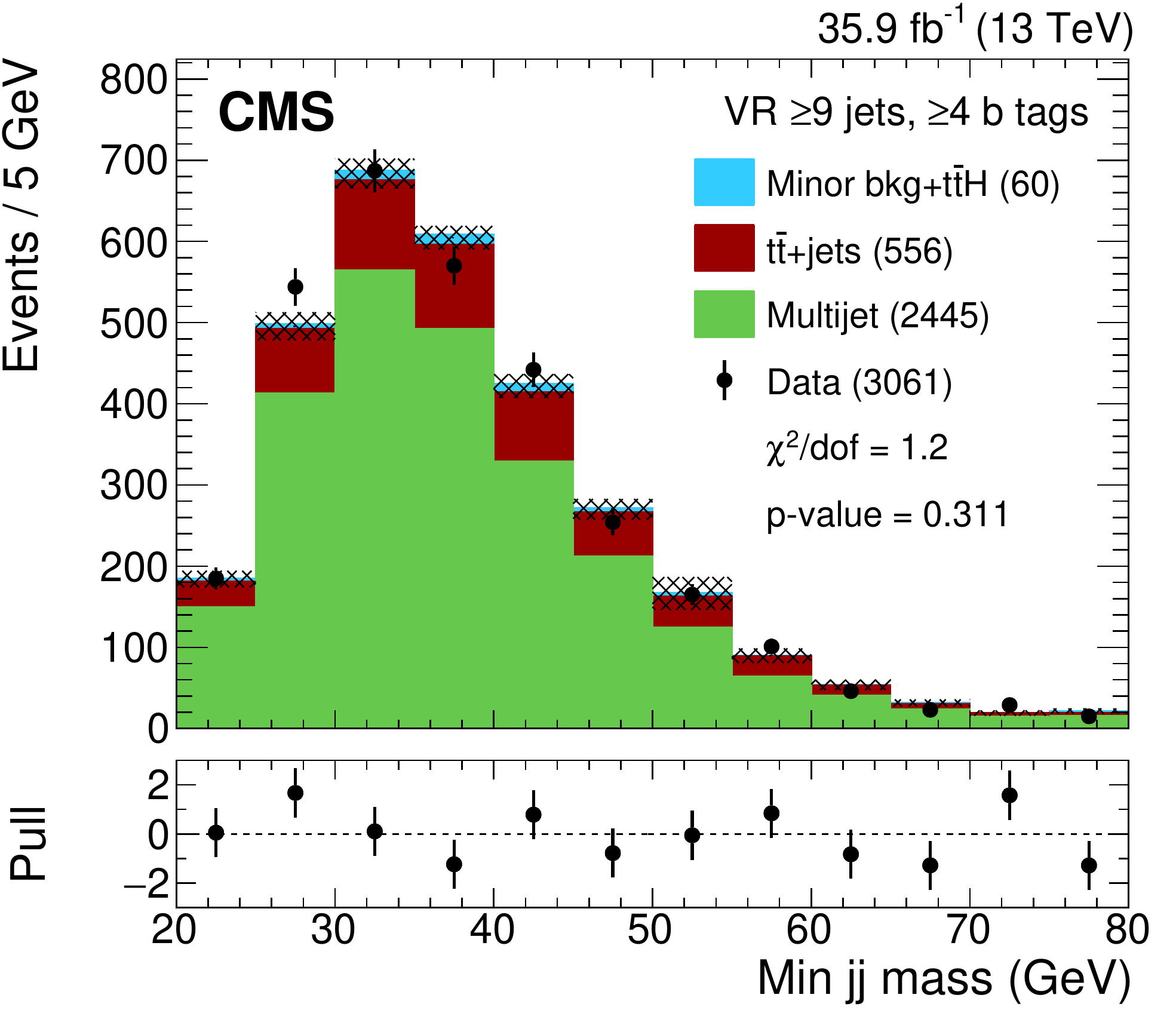} \\
   \vspace{10pt}
   \caption{Distributions in data, in simulated backgrounds, and in the estimated multijet background for the ($\geq$9j,~$\geq$4{\cPqb}) VR category. The MEM discriminant (upper left), \HT (upper right), \pt of the leading jet (middle left), minimum $\Delta R$ between {\cPqb} jets (middle right), invariant mass of the closest {\cPqb} jet pair (lower left), and minimum mass of all jet pairs (lower right). The level of agreement between data and estimation is expressed in terms of a $\chi^2$ divided by the number of degrees of freedom (dof), which are given along with their corresponding statistical $p$-values. The differences between data and the total estimates divided by the quadratic sum of the statistical and systematic uncertainties in the data and in the estimates (pulls) are given below the main panels. The numbers in parenthesis in the legends represent the total yields for the corresponding entries. The last bin includes event overflows.}
   \label{fig:qcd:VR}
\end{figure}

A consistency check of the procedure used to estimate the multijet background is performed through simulation.
Since the power of this test is limited by the statistical uncertainty in simulated QCD multijet events, the method is validated in data using events with $\mathrm{QGLR}<0.5$, by applying the same procedure used to estimate the multijet background in the SR.
The distributions in the MEM discriminant, described in Section~\ref{sec:strategy}, as well as several kinematic variables in the VR in data, together with the multijet background estimate obtained from data,
and other simulated contributions, are shown for the most sensitive event category, ($\geq$9j,~$\geq$4{\cPqb}), in Fig.~\ref{fig:qcd:VR}. All discrepancies are taken into account via the systematic uncertainties in the multijet background, particularly in events with low \HT in the (7j,~3{\cPqb}) category and in all $\geq$4{\cPqb} categories, in which two uncorrelated uncertainties are applied, as discussed in Section~\ref{sec:syst}.

To verify that the good performance demonstrated in the gluon jet enriched VR ($\mathrm{QGLR}<0.5$) also holds in the quark jet enriched SR ($\mathrm{QGLR}>0.5$), we investigate another CR.
Specifically, the {\cPqb} tagging criteria are changed by selecting jets that fulfil some intermediate but not the formal medium {\cPqb} tagging requirement, which corresponds to an efficiency for tagging {\cPqb} jets of $\approx$75\% with a misidentification rate of $\approx$4\%.
The classified jets are then used to form jet and {\cPqb} jet multiplicity categories in analogy with the SR.
Since these categories are orthogonal to the categories in the SR, they are used to verify that the background estimation is valid for $\mathrm{QGLR}>0.5$.
In all validation regions, this background estimation reproduces the kinematic distributions measured in data.
In the remainder of this paper, all multijet estimates are based on data, unless stated to the contrary.

\section{Signal extraction}
\label{sec:strategy}
A likelihood technique based on the LO matrix elements for the \ttH signal and the \ttbb background processes is used to extract the signal. This method utilizes the full kinematic properties
of each event to provide a powerful MEM discriminant between the signal and background. Although the discriminant is constructed to discriminate against the \ttbb background,
it performs well against \ttlight jets and against multijet events. Consequently, it is used as the single discriminant against all background sources.
The MEM algorithm is similar to the one documented in Ref.~\cite{Khachatryan:2015ila}, but adapted to the all-jet final state.

\begin{figure}[tbh]
  \centering
  \raisebox{-0.5\height}{\includegraphics[width=0.6\cmsFigWidth]{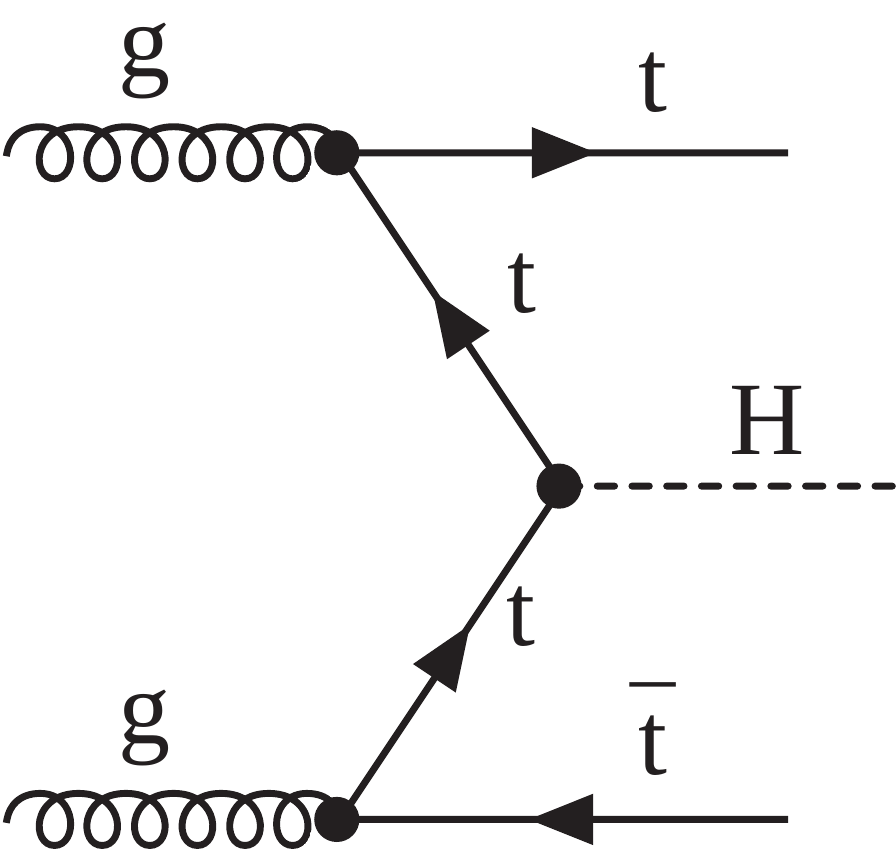}}
   \hspace{5ex}
   \raisebox{-0.5\height}{\includegraphics[width=0.6\cmsFigWidth]{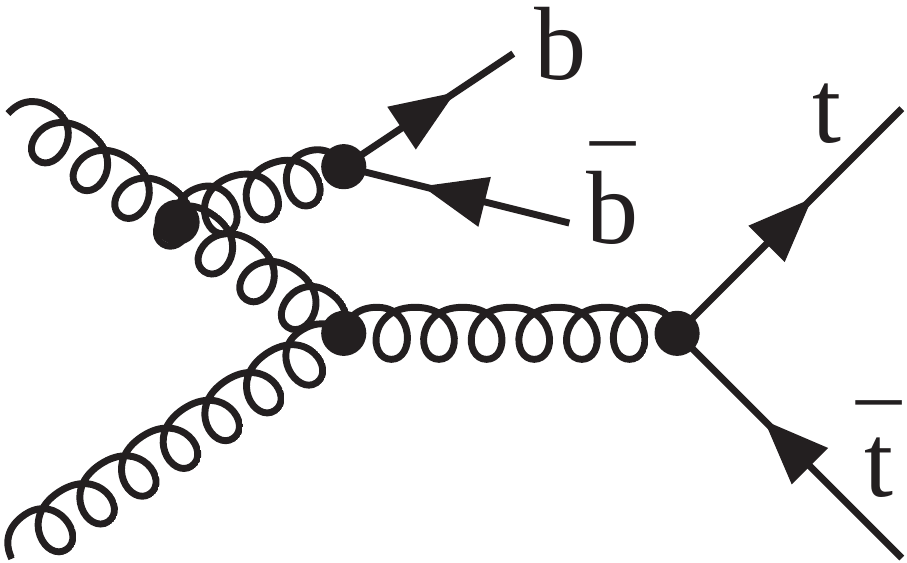}}
   \vspace{5pt}
   \caption{Examples of LO
     Feynman diagrams for the partonic processes of $\cPg\cPg \to \ttH$
     and $\cPg\cPg \to \ttbb$.}
   \label{fig:LO_feynman}
\end{figure}

The MEM probability density functions for signal and background are constructed at LO assuming that both reactions proceed via gluon-gluon fusion, which is valid, as this production mechanism represents most of the event yield. For example, at $\sqrt{s}=14\TeV$, the fraction of gluon-gluon initiated \ttH subprocesses corresponds to $\approx$80\% of the inclusive NLO cross section~\cite{Dawson:2003zu}. Example Feynman diagrams for these processes are shown in Fig.~\ref{fig:LO_feynman}.
All inequivalent jet--quark associations in the reconstruction of the final state are considered in the analysis.
Hadronization and detector effects are taken into account via transfer functions obtained from simulation that map the measured jet four-momenta to the final-state particles.

Each event contains three or four jets that most likely originate from {\cPqb} quarks (according to their CSVv2 discriminant values), and are considered as candidates for {\cPqb} quarks from the \Hbb and $\cPqt \to \PW \cPqb$ decays, whereas untagged jets are considered as candidates for the $\PW \to \cPqu \PAQd$ or $\cPqc \PAQs$ quarks (and their charge conjugates).
The accepted jet permutations must retain the assigned ``\cPqb'' status of their quarks.
To account for the loss of jets because of limited detector acceptance, as well as the presence of additional jets from gluon radiation, the method integrates over certain final-state variables in many categories, specifically over quark directions. In events with eight jets and four {\cPqb} jets, one light-flavour jet is excluded from our MEM calculation in turn, and a sum is taken over the additional permutations. This approach has been checked and shown to provide improved performance relative to the fully reconstructed hypothesis in the (8j,~$\ge$4{\cPqb}) category. Events with only three {\cPqb} jets are assumed to have lost a bottom quark from the decay of a top quark. Up to five untagged jets are considered as other than {\cPqb} quark candidates, while additional untagged jets are ignored. For five such quark candidates, one is excluded in turn and the number of permutations is increased by a factor of five.
The final choice of hypothesis for each category, optimized according to discrimination power and computing performance, is given in Table~\ref{table:mem_hypos}.

\begin{table}[hbtp]
  \centering
    \topcaption{Selected MEM hypotheses
     for each event topology. The 4W2H1T hypothesis assumes 1 {\cPqb} quark from a top quark is lost,  3W2H2T assumes that 1 quark from a {\PW} boson is lost, and 4W2H2T represents the fully reconstructed hypothesis requiring at least 8 jets.
     }
    \begin{tabular}{r r c}
    \hline
    \multicolumn{2}{c}{Category} & MEM hypothesis \\
    \hline
    7 jets, & \hspace{-4.15mm} 3 {\cPqb} jets& 4W2H1T  \\
    8 jets, & \hspace{-4.15mm} 3 {\cPqb} jets   & 4W2H1T  \\
    $\geq$9 jets, & \hspace{-4.15mm} 3 {\cPqb} jets   & 4W2H1T \\
    7 jets, & \hspace{-4.15mm} $\geq$4 {\cPqb} jets   & 3W2H2T \\
    8 jets, & \hspace{-4.15mm} $\geq$4 {\cPqb} jets   & 3W2H2T \\
    $\geq$9 jets, & \hspace{-4.15mm} $\geq$4 {\cPqb} jets  & 4W2H2T \\
    \hline
    \end{tabular}
  \label{table:mem_hypos}
\end{table}

For each event, the MEM probability density function under the signal (S) or background (B) hypothesis, \ie $\mathcal{H} = \ttH$ (S) or $\mathcal{H} = \ttbb$ (B), is calculated as:
\begin{align}\label{eq:p}
w(\mathbf{y}|\mathcal{H}) = \sum_\textrm{perm} \int \rd\mathbf{x}
\int \frac{{\rd}x_a {\rd}x_b}{2 x_a x_b s} g(x_a,Q)g(x_b,Q) \delta^4 \left ( p_a + p_b - \sum_{k=1}^8 p_k \right )
|\mathcal{M}_\mathcal{H}(p_a, p_b, \mathbf{p})|^2 W(\mathbf{y}|\mathbf{p}),
\end{align}
where $\mathbf{y}$ represents the set of measured observables, \ie jet momenta, and $s$ is the square of the {\Pp\Pp} centre-of-mass energy.
The sum is taken over the possible permutations of matching observed jets to final-state quarks.
The integration is performed over the phase space volume, $\mathbf{x}$, of the final-state quarks, with momenta $\mathbf{p} = (p_1, \ldots, p_8)$, and over the incident gluon momentum fractions, $x_a$ and $x_b$, using the \textsc{vegas}~\cite{Lepage:1977sw} algorithm.
The gluon PDF
$g$ is calculated in \textsc{lhapdf}~\cite{Buckley:2014ana} at the factorization scale $Q$, which is the momentum transfer in the event, using the CTEQ6.6 set of PDFs~\cite{Nadolsky:2008zw}.
The four-dimensional delta function is reduced to two dimensions, which ensures conservation of energy and longitudinal momentum, while the total \pt of the final-state quarks is weakly constrained via a resolution function, $\mathcal{R}(\vec{\rho}_\mathrm{T})$,
where $\vec{\rho}_\mathrm{T}$ is the measured transverse recoil, equal to the difference between \ptvecmiss and the vector \pt sum of the specifically selected jets.
The scattering amplitude for the given hypothesis, $\mathcal{M}_\mathcal{H}$, is calculated at LO using \textsc{OpenLoops} (v1.3.1)~\cite{Cascioli:2011va}, and all resonances are treated using the narrow-width approximation~\cite{Kauer:2007zc}. The masses of the Higgs boson, the {\PW} boson, and the top quark are fixed, thereby constraining the energy of certain final-state quarks.
The final integrations of the fully reconstructed \ttH and \ttbb hypotheses are therefore performed over the incoming gluon momentum fractions and three or four respective final-state quark energies, with an integration range of a factor of three that of the associated jet energy resolution (JER).

The transfer function $W(\mathbf{y} | \mathbf{p} )$ corresponds to the likelihood of measuring the set of variables $\mathbf{y}$ given the final-state quark momenta $\mathbf{p}$. Given the excellent angular resolution of reconstructed jets, the direction of the quarks is assumed to be precisely measured, which reduces the total transfer function to a product of quark energy transfer functions. Each quark energy transfer function is modelled via a double Gaussian with different parameterizations for jets associated with {\cPqb} and other quarks, all obtained from MC simulation.

Using Eq.~(\ref{eq:p}), the MEM event discriminant is defined as:
\begin{equation}\label{eq:psb}
\mathcal{P}_{\mathrm{S}/\mathrm{B}} = \frac{w(\mathbf{y}|\mathrm{S})}{w(\mathbf{y}|\mathrm{S}) + \kappa_{\mathrm{S}/\mathrm{B}} w(\mathbf{y}|\mathrm{B})},
\end{equation}
where $\kappa_{\mathrm{S}/\mathrm{B}}$ is a positive constant that adjusts the relative normalization between $w(\mathbf{y}|\mathrm{S})$ and $w(\mathbf{y}|\mathrm{B})$, given that the individual probability densities are not normalized to unity. The $\kappa_{\mathrm{S}/\mathrm{B}}$ constant is optimized category-by-category by minimizing the expected exclusion limit on the signal strength discussed in Section~\ref{sec:result}. Starting with a value that provides good visual discrimination between signal and background, the optimization is performed by testing different values of $\kappa_{\mathrm{S}/\mathrm{B}}$ above and below its original value until a minimum in the expected limit is found in each category as well as in combinations of categories. This optimization was proven to be equivalent to optimizing the significance of discovery.
By construction, the MEM discriminant satisfies the condition $0 \le \mathcal{P}_{\mathrm{S}/\mathrm{B}}  \le 1$.

The event yields expected for the signal and the different background processes, and the yields observed in data, are listed in Table~\ref{table:cats_yields} for each analysis category.
The proportion of expected \ttH events increases with the number of jets and of {\cPqb} jets. To better represent the signal strength in each category, the expected yields for signal and background in the last three bins of the MEM discriminant are listed in Table~\ref{table:cats_yields_bin}.

\begin{table}[hbtp]
  \topcaption{Expected numbers of \ttH signal and background events, and the observed event yields for the six analysis categories, following the fit to data. The signal contributions are given at the best fit value. The quoted uncertainties contain all the contributions described in Section~\ref{sec:syst} added in quadrature, considering all correlations among the processes.
  }
  \label{table:cats_yields}
  \centering
    \renewcommand{\arraystretch}{1.1}
    \resizebox{\textwidth}{!}
    {
    \begin{tabular}{l r l r l r l r l r l r l}
    \hline
    Process & \multicolumn{2}{c}{7j, 3{\cPqb}} & \multicolumn{2}{c}{8j, 3{\cPqb}} & \multicolumn{2}{c}{$\geq$9j, 3{\cPqb}} & \multicolumn{2}{c}{7j, $\geq$4{\cPqb}} & \multicolumn{2}{c}{8j, $\geq$4{\cPqb}} & \multicolumn{2}{c}{$\geq$9j, $\geq$4{\cPqb}} \\
    \hline
Multijet	& 47\,600 & \hspace{-4.15mm} $\pm$ 3\,000	& 32\,700 & \hspace{-4.15mm} $\pm$ 2\,200	& 17\,600 & \hspace{-4.15mm} $\pm$ 1\,600	& 3\,530 & \hspace{-4.15mm} $\pm$ 270	& 3\,770 & \hspace{-4.15mm} $\pm$ 360	& 2\,280 & \hspace{-4.15mm} $\pm$ 290 \\	
\ttlf	& 7\,700 & \hspace{-4.15mm} $\pm$ 1\,600	& 5\,700 & \hspace{-4.15mm} $\pm$ 1\,100	& 3\,160 & \hspace{-4.15mm} $\pm$ 550	& 310 & \hspace{-4.15mm} $\pm$ 130	& 410 & \hspace{-4.15mm} $\pm$ 220	& 244 & \hspace{-4.15mm} $\pm$ 96 \\	
\ttcc	& 3\,100 & \hspace{-4.15mm} $\pm$ 1\,400	& 2\,800 & \hspace{-4.15mm} $\pm$ 1\,200	& 2\,170 & \hspace{-4.15mm} $\pm$ 970	& 190 & \hspace{-4.15mm} $\pm$ 100	& 270 & \hspace{-4.15mm} $\pm$ 150	& 270 & \hspace{-4.15mm} $\pm$ 150 \\	
\ttb	& 1\,400 & \hspace{-4.15mm} $\pm$ 620	& 1\,240 & \hspace{-4.15mm} $\pm$ 620	& 890 & \hspace{-4.15mm} $\pm$ 420	& 142 & \hspace{-4.15mm} $\pm$ 80	& 160 & \hspace{-4.15mm} $\pm$ 110	& 134 & \hspace{-4.15mm} $\pm$ 73 \\	
\tttwob	& 890 & \hspace{-4.15mm} $\pm$ 450	& 760 & \hspace{-4.15mm} $\pm$ 370	& 600 & \hspace{-4.15mm} $\pm$ 290	& 87 & \hspace{-4.15mm} $\pm$ 58	& 114 & \hspace{-4.15mm} $\pm$ 77	& 101 & \hspace{-4.15mm} $\pm$ 52 \\	
\ttbb	& 870 & \hspace{-4.15mm} $\pm$ 340	& 1\,010 & \hspace{-4.15mm} $\pm$ 370	& 970 & \hspace{-4.15mm} $\pm$ 380	& 203 & \hspace{-4.15mm} $\pm$ 90	& 370 & \hspace{-4.15mm} $\pm$ 150	& 410 & \hspace{-4.15mm} $\pm$ 170 \\	
Single {\cPqt}	& 750 & \hspace{-4.15mm} $\pm$ 190	& 520 & \hspace{-4.15mm} $\pm$ 130	& 284 & \hspace{-4.15mm} $\pm$ 75	& 43 & \hspace{-4.15mm} $\pm$ 20	& 78 & \hspace{-4.15mm} $\pm$ 68	& 35 & \hspace{-4.15mm} $\pm$ 17 \\	
\Vjets	& 460 & \hspace{-4.15mm} $\pm$ 170	& 290 & \hspace{-4.15mm} $\pm$ 110	& 240 & \hspace{-4.15mm} $\pm$ 220	& 36 & \hspace{-4.15mm} $\pm$ 33	& 45 & \hspace{-4.15mm} $\pm$ 110	& 17 & \hspace{-4.15mm} $\pm$ 12 \\
\ttV	& 110 & \hspace{-4.15mm} $\pm$ 20	& 122 & \hspace{-4.15mm} $\pm$ 27	& 120 & \hspace{-4.15mm} $\pm$ 30	& 14 & \hspace{-4.15mm} $\pm$ 7	& 28 & \hspace{-4.15mm} $\pm$ 14	& 28 & \hspace{-4.15mm} $\pm$ 14 \\	
Diboson	& 14 & \hspace{-4.15mm} $\pm$ 5	& 5 & \hspace{-4.15mm} $\pm$ 4	& 1 & \hspace{-4.15mm} $\pm$ 1	& 0.6 & \hspace{-4.15mm} $\pm$ 0.5	& 0.6 & \hspace{-4.15mm} $\pm$ 0.6	& 0 & \hspace{-4.15mm} $\pm$ 10 \\[\cmsTabSkip]
Total bkg	& 62\,790 & \hspace{-4.15mm} $\pm$ 900	& 45\,220 & \hspace{-4.15mm} $\pm$ 850	& 26\,020 & \hspace{-4.15mm} $\pm$ 640	& 4\,550 & \hspace{-4.15mm} $\pm$ 180	& 5\,240 & \hspace{-4.15mm} $\pm$ 340	& 3\,520 & \hspace{-4.15mm} $\pm$ 190 \\[\cmsTabSkip]
\ttH ($\muhat=0.9$)	& 130 & \hspace{-4.15mm} $\pm$ 210	& 140 & \hspace{-4.15mm} $\pm$ 220	& 120 & \hspace{-4.15mm} $\pm$ 190	& 32 & \hspace{-4.15mm} $\pm$ 51	& 46 & \hspace{-4.15mm} $\pm$ 75	& 48 & \hspace{-4.15mm} $\pm$ 77 \\	
Data	& \multicolumn{2}{c}{62\,920}	& \multicolumn{2}{c}{45\,359}	& \multicolumn{2}{c}{26\,136}	& \multicolumn{2}{c}{4\,588}	& \multicolumn{2}{c}{5\,287}	& \multicolumn{2}{c}{3\,566} \\		
   \hline
    \end{tabular}
    }
\end{table}

\begin{table}[hbtp]
  \topcaption{Expected numbers of \ttH signal and background events in the last three bins of the MEM discriminant for the six analysis categories, following the fit to data. The signal contributions are given at the best fit value. The quoted uncertainties contain all the contributions described in Section~\ref{sec:syst} added in quadrature, considering all correlations among the processes.
   Also given are the signal (S) to total background (B) ratios for the SM \ttH expectation ($\mu = 1$).}
  \label{table:cats_yields_bin}
  \centering
    \renewcommand{\arraystretch}{1.1}
    \resizebox{\textwidth}{!}
    {
    \begin{tabular}{l r l r l r l r l r l r l}
    \hline
    Process & \multicolumn{2}{c}{7j, 3{\cPqb}} & \multicolumn{2}{c}{8j, 3{\cPqb}} & \multicolumn{2}{c}{$\geq$9j, 3{\cPqb}} & \multicolumn{2}{c}{7j, $\geq$4{\cPqb}} & \multicolumn{2}{c}{8j, $\geq$4{\cPqb}} & \multicolumn{2}{c}{$\geq$9j, $\geq$4{\cPqb}} \\
    \hline
Multijet	& 8\,560 & \hspace{-4.15mm} $\pm$ 820	& 5\,510 & \hspace{-4.15mm} $\pm$ 590	& 3\,120 & \hspace{-4.15mm} $\pm$ 400	& 608 & \hspace{-4.15mm} $\pm$ 61	& 748 & \hspace{-4.15mm} $\pm$ 74	& 376 & \hspace{-4.15mm} $\pm$ 57 \\	
\ttlf	& 3\,300 & \hspace{-4.15mm} $\pm$ 470	& 2\,220 & \hspace{-4.15mm} $\pm$ 330	& 1\,110 & \hspace{-4.15mm} $\pm$ 170	& 121 & \hspace{-4.15mm} $\pm$ 34	& 107 & \hspace{-4.15mm} $\pm$ 31	& 54 & \hspace{-4.15mm} $\pm$ 21 \\	
\ttcc	& 1\,050 & \hspace{-4.15mm} $\pm$ 410	& 920 & \hspace{-4.15mm} $\pm$ 370	& 660 & \hspace{-4.15mm} $\pm$ 260	& 53 & \hspace{-4.15mm} $\pm$ 27	& 88 & \hspace{-4.15mm} $\pm$ 39	& 70 & \hspace{-4.15mm} $\pm$ 30 \\	
\ttb	& 380 & \hspace{-4.15mm} $\pm$ 160	& 330 & \hspace{-4.15mm} $\pm$ 140	& 240 & \hspace{-4.15mm} $\pm$ 110	& 44 & \hspace{-4.15mm} $\pm$ 27	& 51 & \hspace{-4.15mm} $\pm$ 25	& 33 & \hspace{-4.15mm} $\pm$ 17 \\	
\tttwob	& 208 & \hspace{-4.15mm} $\pm$ 94	& 167 & \hspace{-4.15mm} $\pm$ 79	& 128 & \hspace{-4.15mm} $\pm$ 59	& 21 & \hspace{-4.15mm} $\pm$ 12	& 36 & \hspace{-4.15mm} $\pm$ 22	& 15 & \hspace{-4.15mm} $\pm$ 8 \\	
\ttbb	& 192 & \hspace{-4.15mm} $\pm$ 64	& 249 & \hspace{-4.15mm} $\pm$ 84	& 228 & \hspace{-4.15mm} $\pm$ 80	& 46 & \hspace{-4.15mm} $\pm$ 18	& 88 & \hspace{-4.15mm} $\pm$ 30	& 95 & \hspace{-4.15mm} $\pm$ 33 \\	
Single {\cPqt}	& 130 & \hspace{-4.15mm} $\pm$ 32	& 100 & \hspace{-4.15mm} $\pm$ 32	& 50 & \hspace{-4.15mm} $\pm$ 15	& 8 & \hspace{-4.15mm} $\pm$ 5	& 24 & \hspace{-4.15mm} $\pm$ 23	& 5 & \hspace{-4.15mm} $\pm$ 4 \\	
\Vjets	& 79 & \hspace{-4.15mm} $\pm$ 48	& 39 & \hspace{-4.15mm} $\pm$ 20	& 28 & \hspace{-4.15mm} $\pm$ 21	& 1 & \hspace{-4.15mm} $\pm$ 2	& 4 & \hspace{-4.15mm} $\pm$ 5	& 0 & \hspace{-4.15mm} $\pm$ 1 \\
\ttV	& 34 & \hspace{-4.15mm} $\pm$ 6	& 39 & \hspace{-4.15mm} $\pm$ 8	& 32 & \hspace{-4.15mm} $\pm$ 8	& 5 & \hspace{-4.15mm} $\pm$ 3	& 8 & \hspace{-4.15mm} $\pm$ 3	& 7 & \hspace{-4.15mm} $\pm$ 3 \\	
Diboson	& 1 & \hspace{-4.15mm} $\pm$ 1	& 0.24 & \hspace{-4.15mm} $\pm$ 0.23	& 0.00 & \hspace{-4.15mm} $\pm$ 0.00	& 0.33 & \hspace{-4.15mm} $\pm$ 0.29	& 0.00 & \hspace{-4.15mm} $\pm$ 0.00	& 0.13 & \hspace{-4.15mm} $\pm$ 0.23 \\[\cmsTabSkip]	
Total bkg	& 13\,930 & \hspace{-4.15mm} $\pm$ 260 & 9\,580 & \hspace{-4.15mm} $\pm$ 250	& 5\,610 & \hspace{-4.15mm} $\pm$ 170	& 910 & \hspace{-4.15mm} $\pm$ 50	& 1\,154 & \hspace{-4.15mm} $\pm$ 76	& 656 & \hspace{-4.15mm} $\pm$ 44 \\[\cmsTabSkip]	
\ttH ($\muhat=0.9$)	& 56 & \hspace{-4.15mm} $\pm$ 83	& 58 & \hspace{-4.15mm} $\pm$ 89	& 47 & \hspace{-4.15mm} $\pm$ 71	& 18 & \hspace{-4.15mm} $\pm$ 27	& 26 & \hspace{-4.15mm} $\pm$ 38	& 21 & \hspace{-4.15mm} $\pm$ 31 \\	
Data	& \multicolumn{2}{c}{13\,937}	& \multicolumn{2}{c}{9\,620}	& \multicolumn{2}{c}{5\,640}	& \multicolumn{2}{c}{958}	& \multicolumn{2}{c}{1\,162}	& \multicolumn{2}{c}{660} \\
$\text{S}/\text{B}$ ($\mu=1$)	& \multicolumn{2}{c}{0.005}	& \multicolumn{2}{c}{0.007}	& \multicolumn{2}{c}{0.009}	& \multicolumn{2}{c}{0.023}	& \multicolumn{2}{c}{0.025}	& \multicolumn{2}{c}{0.036} \\		
   \hline
    \end{tabular}
    }
\end{table}

\section{Systematic uncertainties}
\label{sec:syst}
Each source of systematic uncertainty is associated with a ``nuisance'' parameter that modifies the likelihood in the final fit discussed in Section~\ref{sec:result}, and can either affect the yield from a process (rate uncertainty), the distribution in the MEM discriminant, or both. In the latter case, the effects from the rate and distribution are treated simultaneously, and can be considered completely correlated. Each individual source of systematic uncertainty is independent of other sources, and its effect on signal and background is completely correlated across the relevant processes.

The uncertainty in the integrated luminosity is estimated to be 2.5\%~\cite{CMS-PAS-LUM-17-001}.
The uncertainties in trigger scale factors are determined from the bin-by-bin uncertainties in the ratio of efficiency in data relative to simulation, and are $\approx$1.0\% on average, with some being as large as 15\%.
The uncertainty in the distribution in the number of pileup interactions is evaluated by changing the total inelastic cross section in the MC simulation by its uncertainty of 4.6\%.
The changes in the weight factors are propagated to the MEM discriminant distributions and treated as fully correlated among all processes.
The impact of the uncertainty in the jet energy scale (JES)~\cite{Chatrchyan:2011ds} is evaluated for each jet in the simulated events by changing the correction factors by their uncertainties, and propagating the effect to the MEM discriminant by recalculating all kinematic quantities. Various independent sources contribute to the overall JES uncertainty, and their impact is therefore evaluated separately and they are treated as uncorrelated in the final fit.
A similar procedure is applied to account for the uncertainty related to the JER, which ranges between about 1 and 5\% of the expected energy resolution, depending on the jet direction. Since the analysis categories are defined in terms of jet multiplicity and kinematics, a change in either JES or JER can induce a migration of events between categories, as well as in the SR. The fractional changes in  event yields induced by one standard deviation shifts in JES or JER range, respectively, from 3--11\% and 2--11\%, depending on the process and the category.

The scale factors applied to correct the CSVv2 discriminant, described in Section~\ref{sec:selection}, are affected by several components of systematic uncertainty, which are attributed to three main sources: JES, purity of heavy- or light-flavour jets in the control sample used to obtain the scale factors, and the statistical uncertainty in the event sample used in their extraction~\cite{CMS-BTV-16-002}. A separate, large uncertainty is applied to charm jets, owing to the lack of a reliable calibration based on data. Each component of these systematic {\cPqb} tagging uncertainties is considered as a separate nuisance parameter in the final fit.

The systematic uncertainties from reweighting the simulated QGL and top quark \pt distributions affect both the rate and the distribution in the MEM discriminant, and are taken into account by changing the functions used in the reweighting procedure by their uncertainties.
Many uncertainties that would normally be related to the MC simulation of the multijet background are avoided by estimating this contribution from data.
Nevertheless, a few small systematic uncertainties remain: an uncertainty in the correction applied to CSVL jets that is estimated
via an alternative correction obtained by applying a second $\eta$ correction; a 2.5 or 2.0\%
uncertainty in the first bin of the distribution in the MEM discriminant across $\ge$4{\cPqb} and 3{\cPqb} categories, applied to account for respective systematic over- or underestimation of this bin in the VR;
a reweighting based on the \HT distribution (considering the \pt of just the first six jets) in the (7j, 3{\cPqb}) and all $\ge$4{\cPqb} categories of the VR, to account for mismodelling at low \HT; and the total normalization in each category that is left unconstrained in the final fit. The uncertainties in the multijet background normalization have the largest impact on the sensitivity of the analysis, and setting the normalization to a fixed value in each category improves the expected limit by 20 to 30\%.

Theoretical uncertainties are treated as process-specific if they only impact a single simulated event sample, otherwise they are treated as correlated across all samples to which they apply. For example, the PDF uncertainty for quark-quark initiated processes, such as \Vjets, \ttW, and diboson events, are treated as 100\% correlated.
The \ttheavy processes represent important sources of irreducible background, which have not yet been reliably measured, nor subjected to higher-order calculations that constrain these contributions. In fact, the most recent direct measurement of the \ttbb cross section in the dilepton final state has an accuracy of $\approx$35\%~\cite{Sirunyan:2017snr}. A 50\% uncertainty in the production rate is therefore assigned separately to the \ttbb, \tttwob, \ttb, and \ttcc processes. These uncertainties have a significant impact on the expected limits, and ignoring these improves the overall limit by about 5\%.

Uncertainties from factorization ($\mu_\text{F}$) and renormalization ($\mu_\text{R}$) scales in the inclusive cross sections range from 1 to 13\%, depending on the process. The PDF uncertainties~\cite{Butterworth:2015oua} range between 2 and 4\%, and are treated as fully correlated for all processes that share the same dominant initial state (\ie $\cPg\cPg$, $\cPg\cPq$, or $\cPq\cPq$), except for \ttH, which is considered separately.
Finally, the limited number of simulated background and signal events leads to statistical fluctuations in the nominal predictions. This is taken into account by assigning a nuisance parameter for each bin of each sample that can be changed by its uncertainty as described in Refs.~\cite{Barlow:1993dm, Conway:2011in}.
Table~\ref{table:systSummary} summarizes the various sources of systematic uncertainty and their impact on yields.

\begin{table}[hbtp] \small
  \topcaption{Summary of the systematic uncertainties affecting the signal and background expectations. The second column indicates the range in yield of the affected processes, caused by changing the nuisance parameters by their uncertainties. The third column indicates whether the uncertainties impact  the distribution in the MEM discriminant. A check mark ($\checkmark$) indicates that the uncertainty applies to the stated processes. An asterisk (*) indicates that the uncertainty affects the multijet distribution indirectly, because of the subtraction of directly affected backgrounds in the CR in data. }
  \label{table:systSummary}
  \centering
  \renewcommand{\arraystretch}{1.1}
  \newcommand{\rowgroup}[1]{\hspace{-1ex}#1}
  \begin{tabular}{l l c c c c c }
    \hline
    \rowgroup{Source}             & Range of & Distribution & \multicolumn{4}{c}{Process} \\
    \cline{4-7}
     & uncertainty (\%) & & \ttH & Multijet & \ttjets & Other \\
    \hline
    \multicolumn{7}{l}{\rowgroup{Experimental uncertainties}} \\
    Integrated luminosity    & 2.5      & No & \checkmark & * & \checkmark & \checkmark \\
    Trigger efficiency          & 1--2    & Yes & \checkmark & * & \checkmark & \checkmark \\
    Pileup                           & 0.2--5 & Yes & \checkmark & * & \checkmark & \checkmark \\
    JES                               & 3--11  & Yes & \checkmark & * & \checkmark & \checkmark \\
    JER                               & 2--11  & Yes & \checkmark & * & \checkmark & \checkmark \\
    {\cPqb} tagging                       & 4--40   & Yes & \checkmark & * & \checkmark & \checkmark \\
    QGL reweighting           & 4--11   & Yes & \checkmark & * & \checkmark & \checkmark \\
    Top quark \pt reweighting   & 1--2   & Yes & \NA & * & \checkmark & \NA \\[\cmsTabSkip]
    \multicolumn{7}{l}{\rowgroup{Multijet estimation}} \\
    CSVL correction  & \NA & Yes & \NA & \checkmark & \NA & \NA \\
    MEM first bin      & \NA & Yes & \NA & \checkmark & \NA & \NA \\
    \HT reweighting & \NA & Yes & \NA & \checkmark & \NA & \NA \\
    Normalization & $\infty$ & No & \NA & \checkmark & \NA & \NA \\[\cmsTabSkip]
    \multicolumn{7}{l}{\rowgroup{Theoretical uncertainties}} \\
    \ttbb normalization & 50 & No & \NA & * & \checkmark & \NA \\
    \tttwob normalization & 50 & No & \NA & * & \checkmark & \NA \\
    \ttb normalization & 50 & No & \NA & * & \checkmark & \NA \\
    \ttcc normalization & 50 & No & \NA & * & \checkmark & \NA \\
    $\mu_\text{F}$/$\mu_\text{R}$ scales for signal           & 6--9 & No & \checkmark & \NA & \NA & \NA \\
    $\mu_\text{F}$/$\mu_\text{R}$ scales for background    & 1--13 & No & \NA & * & \checkmark & \checkmark \\
    PDFs                    & 2--4 & No & \checkmark & * & \checkmark & \checkmark \\
    Simulated sample size  & 2--40 & Yes & \checkmark & * & \checkmark & \checkmark \\
    \hline
  \end{tabular}
\end{table}

\section{Results}
\label{sec:result}
The MEM discriminant is used to extract the signal. The results are interpreted in terms of the signal strength modifier, defined as the ratio of the measured \ttH production cross section to the SM expectation for $\massHiggs = 125 \GeV$ ($\mu = \sigma / \sigma_\text{SM}$). The statistical method used to determine the signal strength is documented in Ref.~\cite{CMS-NOTE-2011-005}. In this method, a likelihood function $\mathcal{L}(\mu, \theta)$ is constructed from the product of Poisson likelihoods for each bin of the MEM discriminant distribution for all six event categories, multiplied by a probability density function for the nuisance parameters $\theta$, discussed in Section~\ref{sec:syst}. The following test statistic, based on the LR, is used to compare competing hypotheses for the value of $\mu$:
\begin{equation}\label{eq:teststat}
q(\mu) = -2 \ln \frac{\mathcal{L}(\mu, \hat{\theta}_\mu)}{\mathcal{L}(\muhat, \hat{\theta})},
\end{equation}
where $\hat{\theta}$ and $\hat{\theta}_\mu$ represent the best fit estimate of $\theta$ for a floating or fixed $\mu$, respectively, following the procedure described in Section 3.2 of Ref.~\cite{Khachatryan:2014jba}.

\begin{figure}[htbp]
  \centering
   \includegraphics[width=\cmsFigWidth]{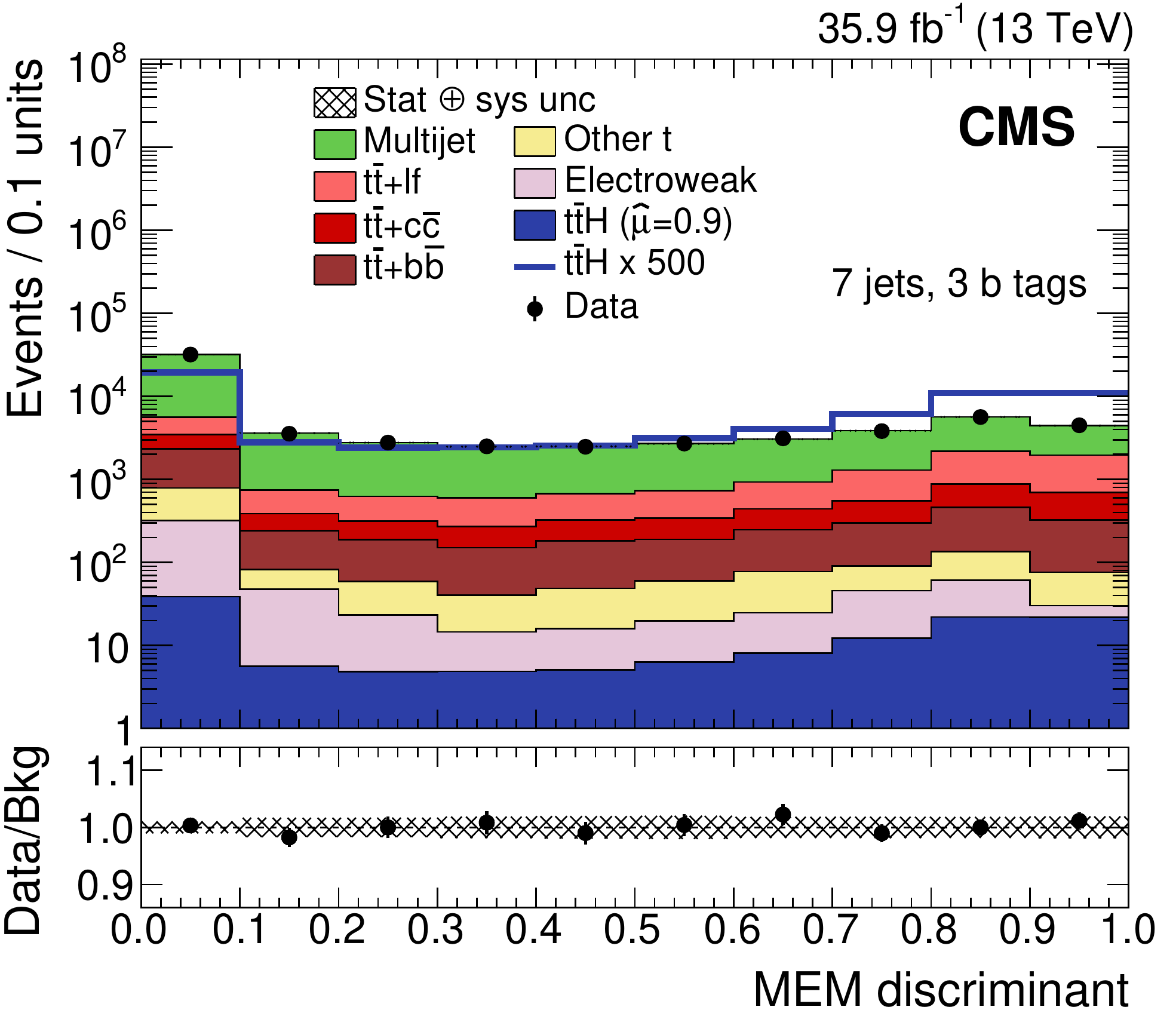}
    \hspace{20pt}
    \includegraphics[width=\cmsFigWidth]{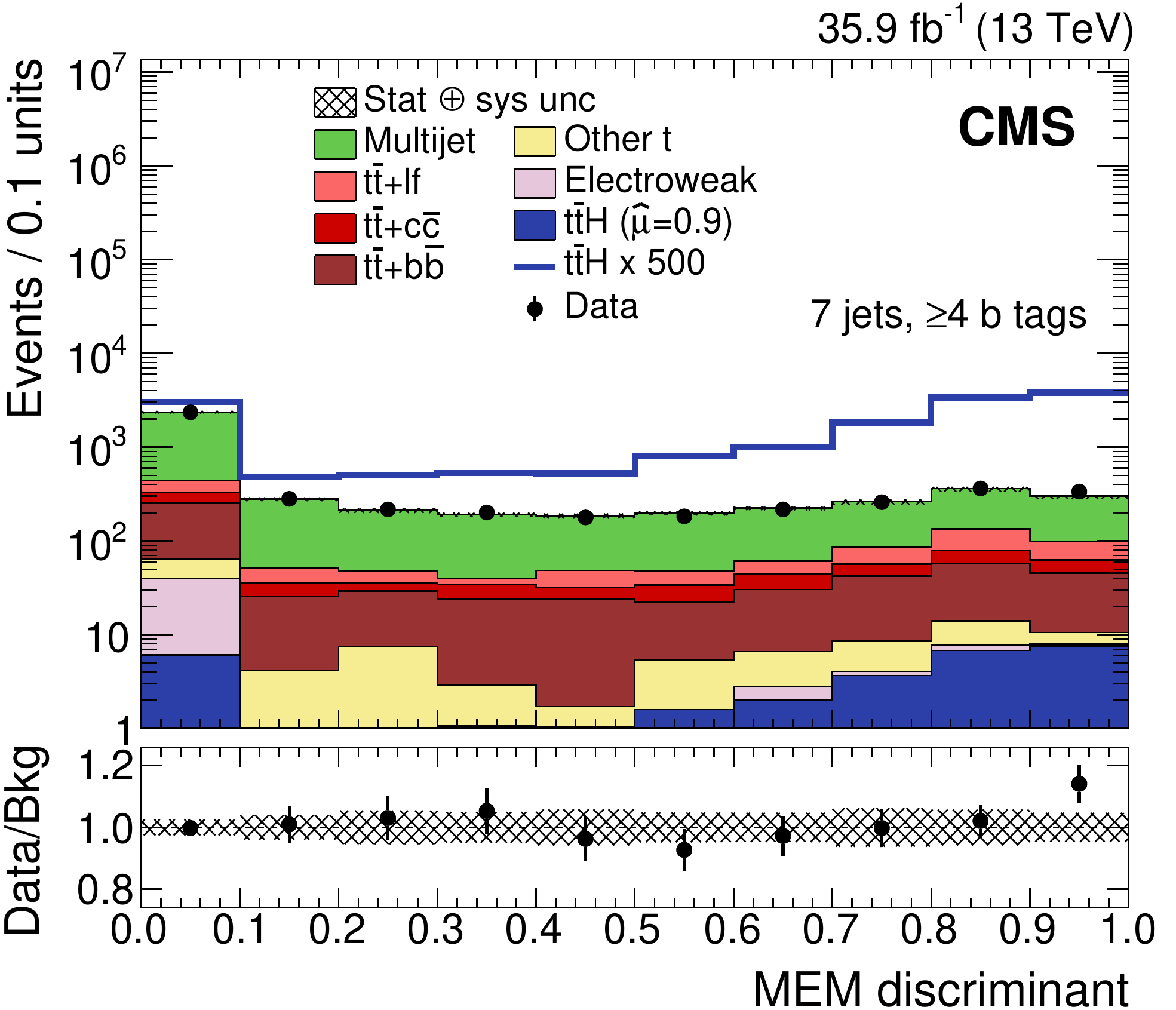} \\
    \vspace{10pt}
    \includegraphics[width=\cmsFigWidth]{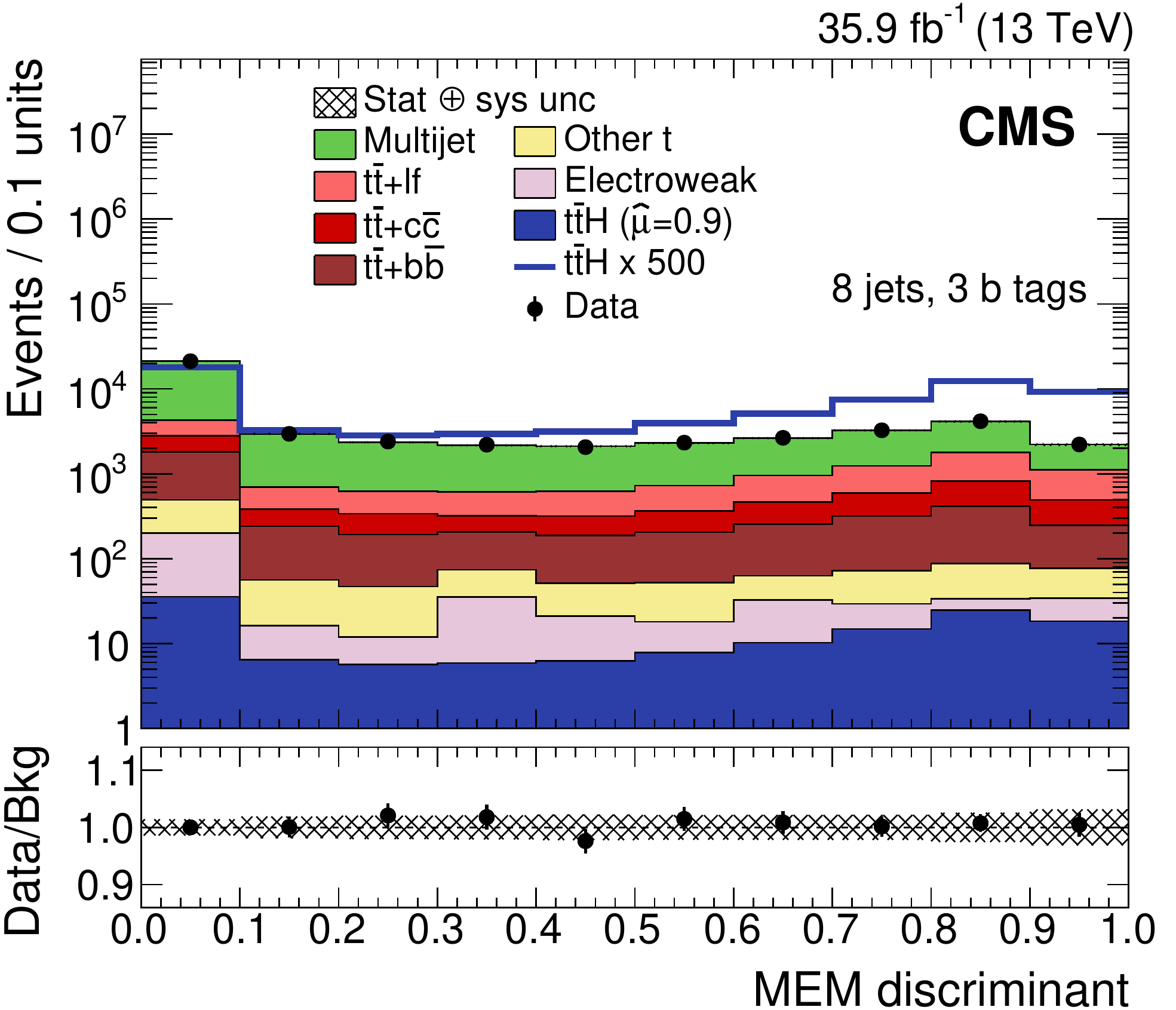}
    \hspace{20pt}
    \includegraphics[width=\cmsFigWidth]{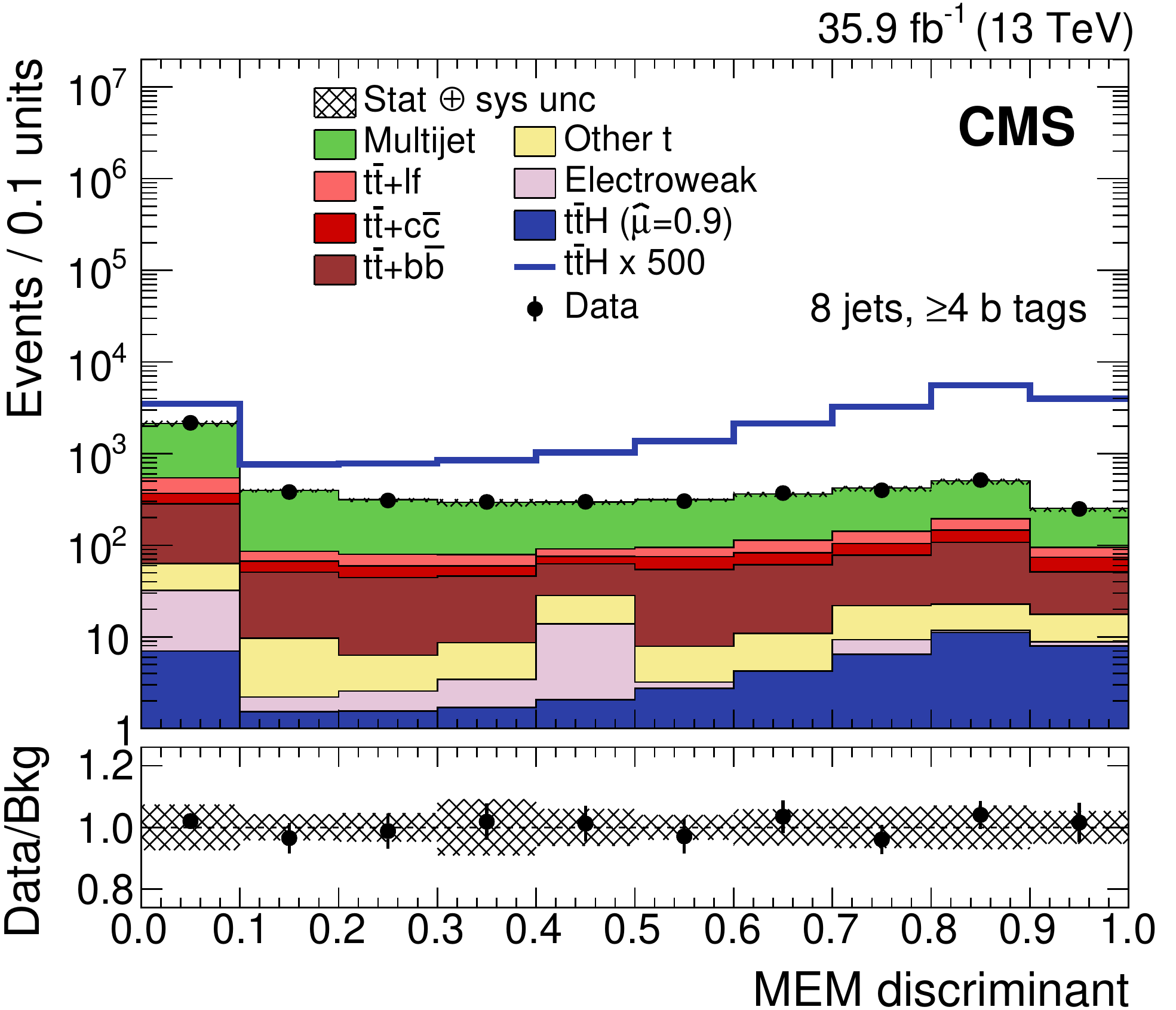} \\
    \vspace{10pt}
    \includegraphics[width=\cmsFigWidth]{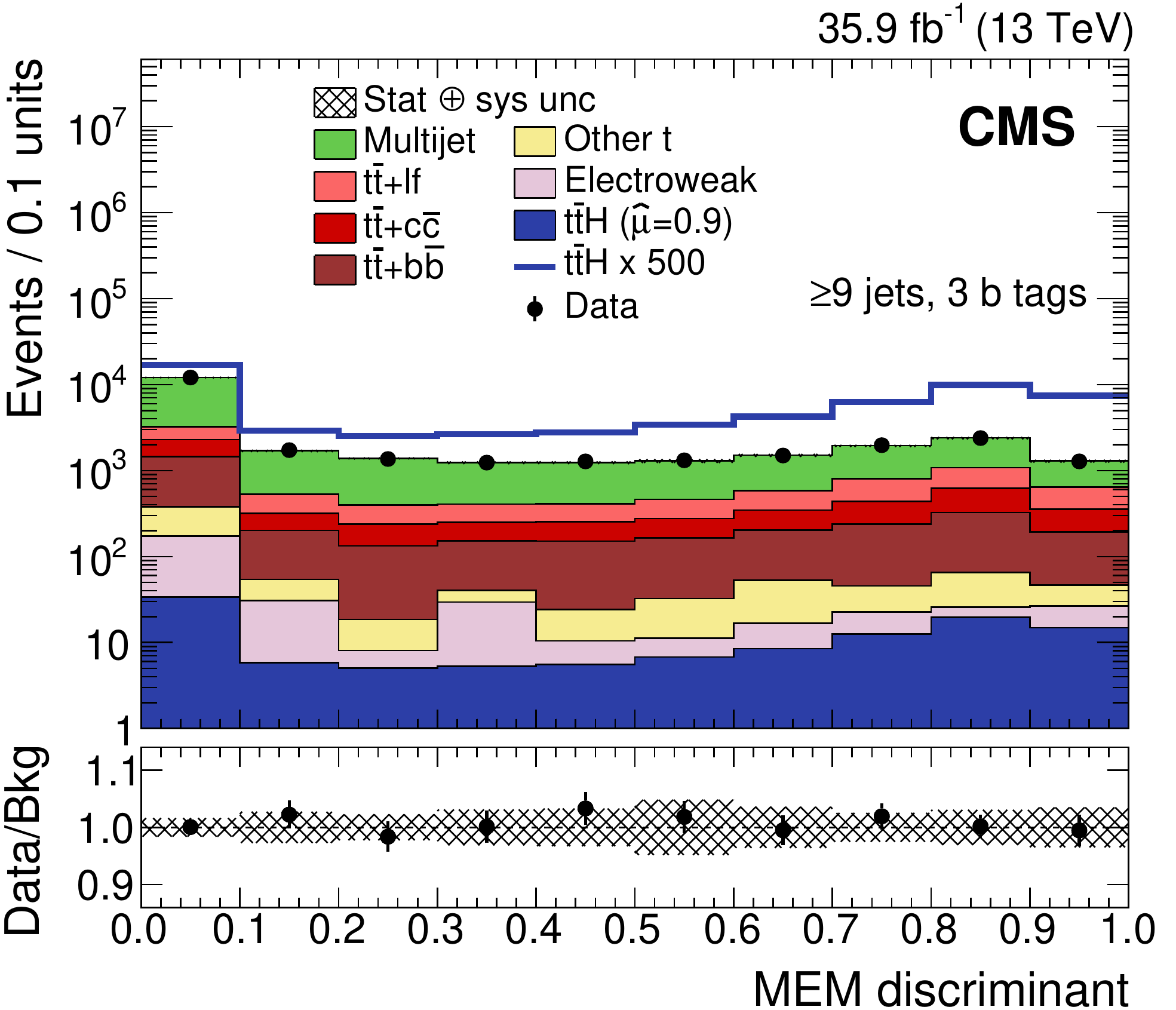}
    \hspace{20pt}
    \includegraphics[width=\cmsFigWidth]{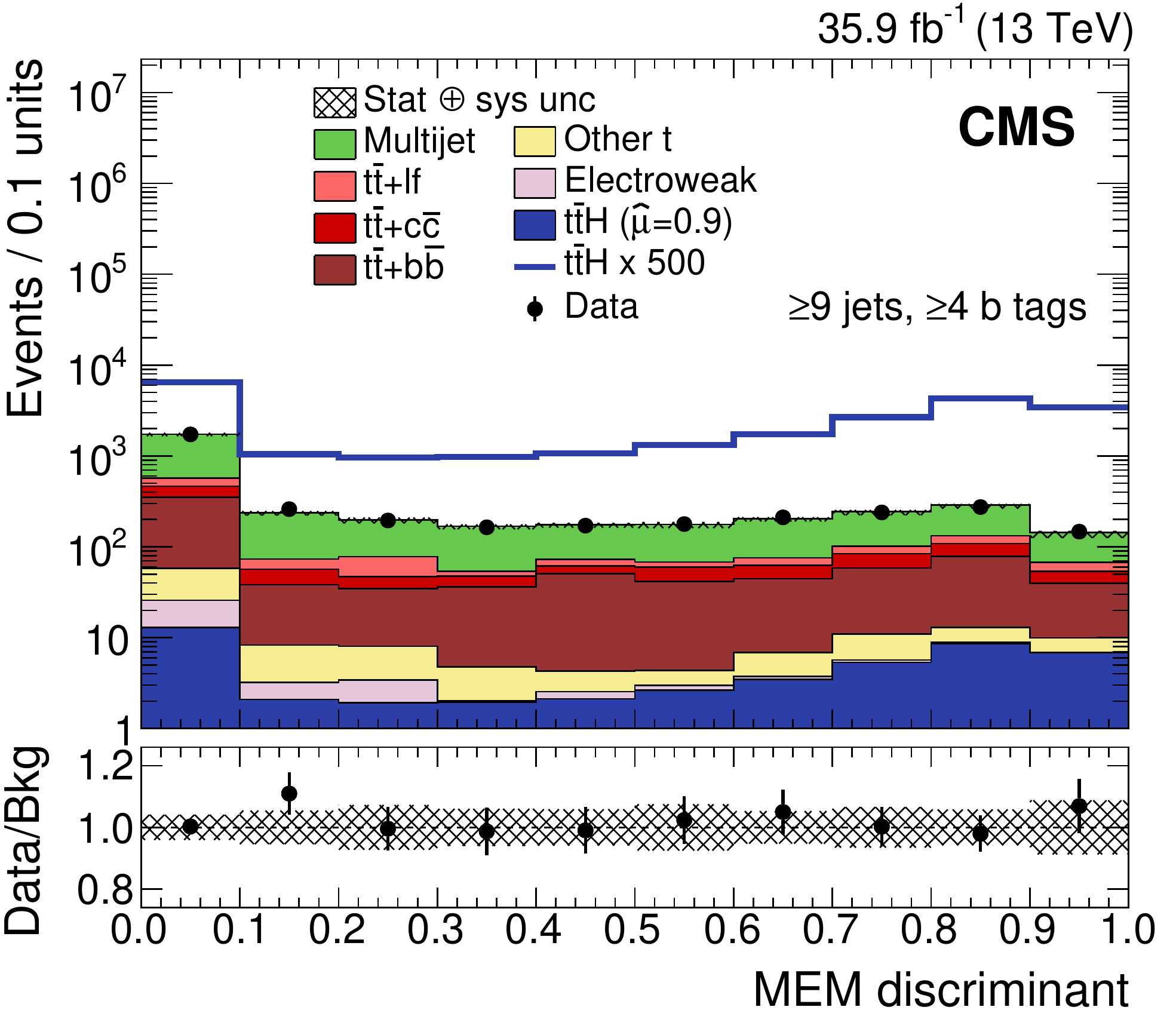} \\
    \vspace{10pt}
   \caption{Distributions in the fitted MEM discriminant for each analysis category.
    The contributions expected from signal and background processes (filled histograms) are shown stacked. The \ttb and \tttwob background process are shown combined with \ttbb, the single {\cPqt} and \ttV processes are shown combined as ``Other {\cPqt}'', and the \Vjets and diboson processes are shown combined as ``Electroweak''.
    The signal distributions (lines) for a Higgs boson mass of $\massHiggs = 125 \GeV$ are multiplied by a factor of 500 and superimposed on the data.
    The distributions in data are indicated by the black points.
    The ratios of data to background appear below the main panels.}
  \label{fig:mem_postfit}
\end{figure}

The final MEM discriminants in each category, following the combined fit to data, are displayed in Fig.~\ref{fig:mem_postfit}. Each background contribution is initially normalized to an integrated luminosity of \lumifh, while the multijet contribution is free to float in the fit. The fit to the nuisance parameters constrains many uncertainties including the multijet normalization. The total uncertainty contains the correlations among the fitted parameters that are sampled through the covariance matrix.

The observed distributions are compared to the signal+background expectation to extract a best fit value of $\muhat = 0.9 \pm 0.7\stat \pm 1.3\syst = 0.9 \pm 1.5 \,(\text{tot})$, which is consistent with the SM expectation of $\mu = 1$.
The statistical component of the total uncertainty is estimated by including in the fit only a limited set of nuisance parameters that are statistical in nature, namely the per-category multijet normalizations and the multijet bin-by-bin uncertainties, which are dominated by the limited event count in the CR. The systematic component is then calculated as the difference in quadrature between the total uncertainty and the statistical component.

\begin{figure}[h!t]
  \centering
  \includegraphics[width=0.49\textwidth]{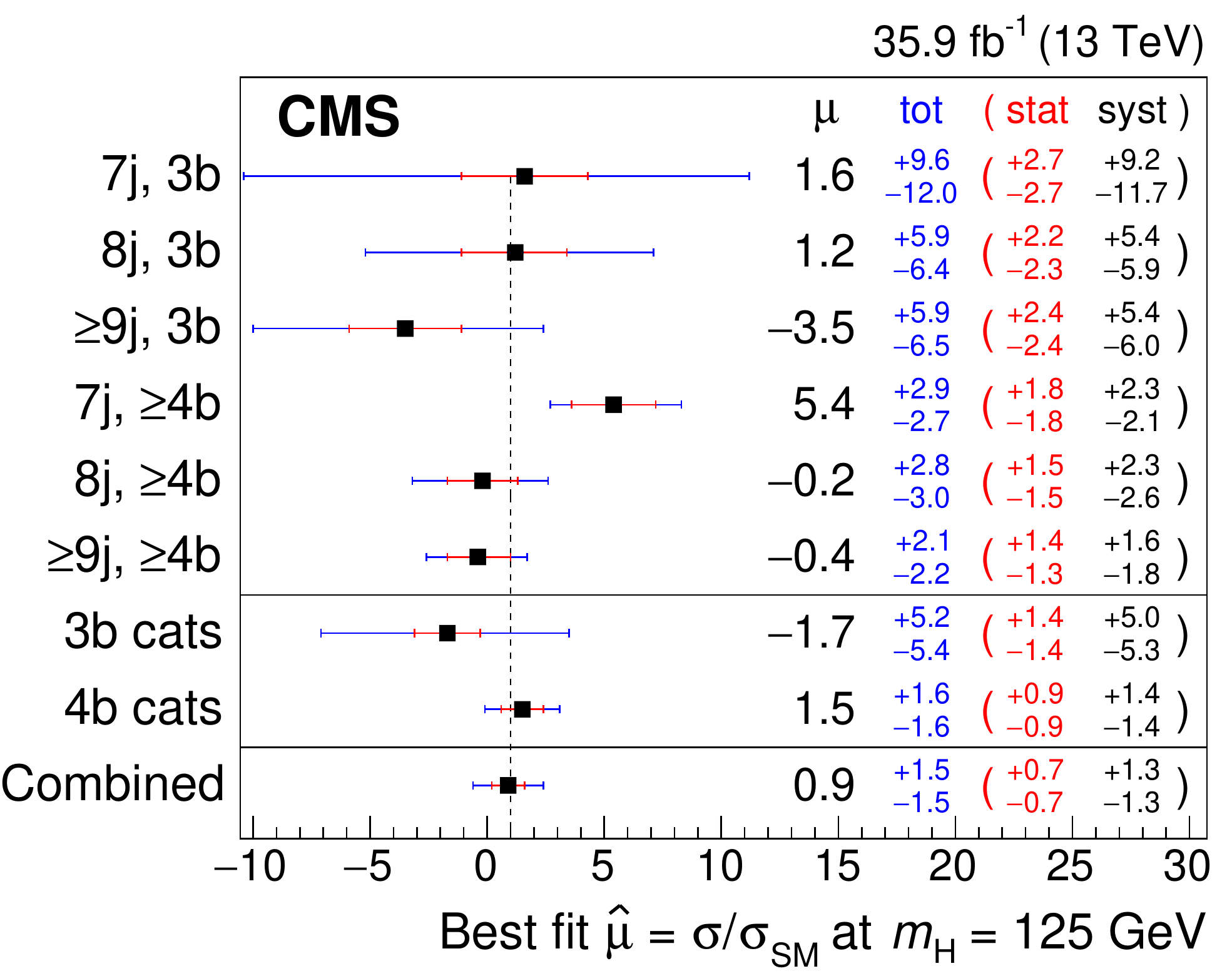}
  \includegraphics[width=0.49\textwidth]{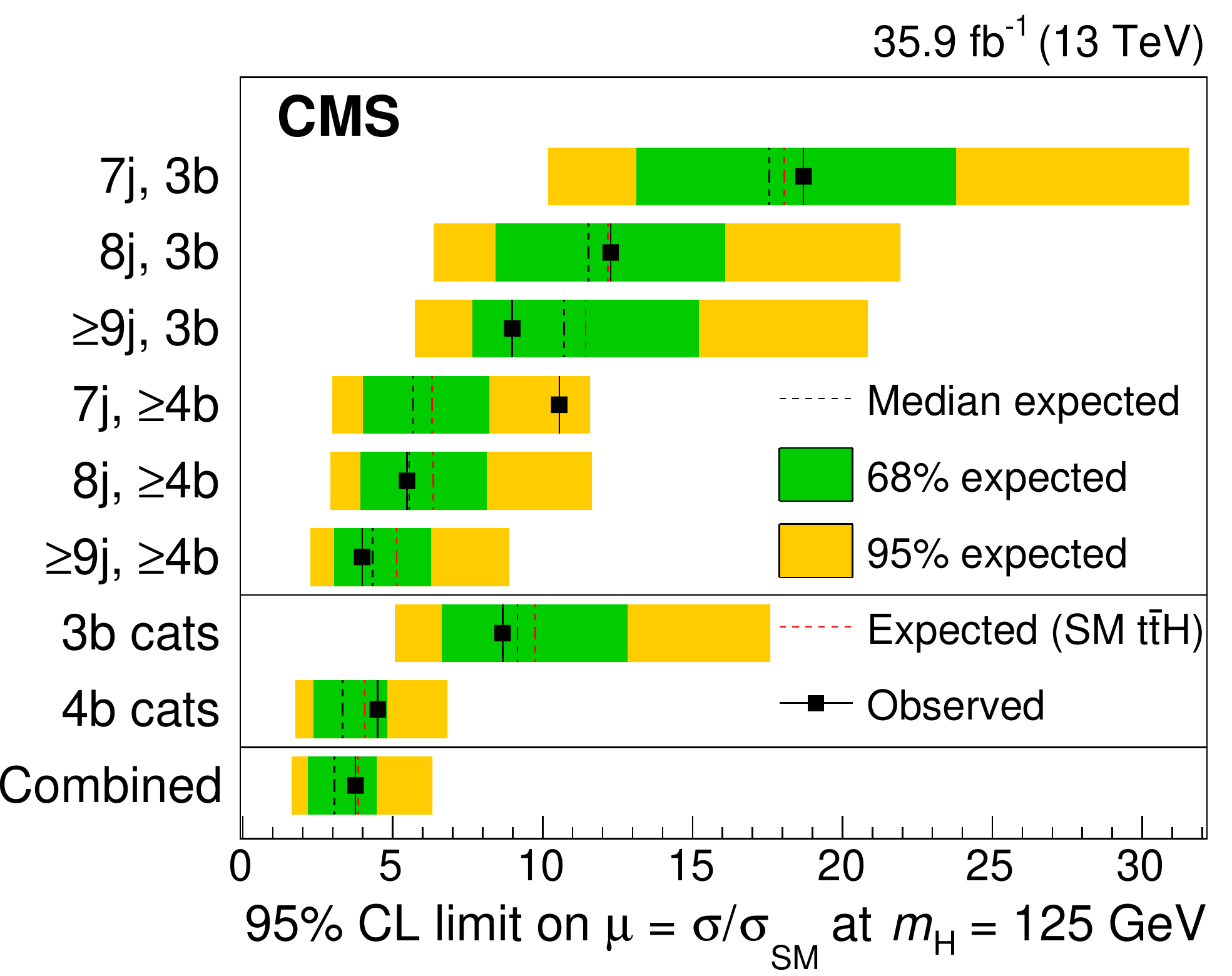}
  \caption{
  Best fit values in the signal strength modifiers $\muhat$, and their 68\% \CL intervals as split into the statistical and systematic components (left), and median expected and observed 95\% \CL upper limits on $\mu$ (right). The expected limits are displayed with their 68\% and 95\% \CL intervals, as well as with the expectation for an injected SM signal of $\mu = 1$.
  }
\label{fig:limit}
\end{figure}

The best fit values of $\mu$ in each analysis category and in the combination are listed in Table~\ref{tab:limit}, and displayed in Fig.~\ref{fig:limit} (left).
Since the value is also compatible with the background-only hypothesis, an exclusion limit at the 95\% \CL can be set using the asymptotic approximation~\cite{Cowan:2010js} of the modified frequentist CL$_\text{s}$ method~\cite{Junk:1999kv, Read:2002hq}.
Combining all categories, we obtain the observed and expected upper limits of $\mu < 3.8$ and $< 3.1$, respectively.
These limits in each category, as well as the combined limit, are listed in Table~\ref{tab:limit} and displayed in Fig.~\ref{fig:limit} (right).

\begin{table}[htbp]
  \topcaption{Best fit value of the signal-strength modifier $\muhat$ and the median
    expected and observed 95\% \CL upper limits (UL) on $\mu$ in each of the six analysis categories, as well as the combined results.
    The best fit values are shown with their total uncertainties and the breakdown into the statistical and systematic components.
    The expected limits are given together with their 68\% \CL intervals.
    }
  \label{tab:limit}
  \centering
    \renewcommand{\arraystretch}{1.8}
    \begin{tabular}{lcrrccccc}
      \hline
      \multirow{2}{*}{Category}& & \multicolumn{3}{c}{Best fit $\muhat$ and uncertainty} & & Observed & & Expected \\[-3.5mm]
          & & $\muhat$ & total & (stat~~syst) & & UL & & UL \\
      \hline
    7j, 3{\cPqb}                        & & $1.6 $ & ${}^{+9.6}_{-12.0}$ & $\bigl({}^{+2.7}_{-2.7}~~{}^{+9.2}_{-11.7}\bigr)$  & & 18.7  & & $17.6^{+6.2}_{-4.4}$ \\
    8j, 3{\cPqb}                         & & $1.2 $ & ${}^{+5.9}_{-6.4}$ & $\bigl({}^{+2.2}_{-2.3}~~{}^{+5.4}_{-5.9}\bigr)$  & & 12.3  & & $11.5^{+4.6}_{-3.1}$ \\
    $\geq$9j, 3{\cPqb}             & & $-3.5 $ & ${}^{+5.9}_{-6.5}$ & $\bigl({}^{+2.4}_{-2.4}~~{}^{+5.4}_{-6.0}\bigr)$  & & 9.0  & & $10.7^{+4.5}_{-3.1}$ \\
    7j,  $\geq$4{\cPqb}            & & $5.4 $ & ${}^{+2.9}_{-2.7}$ & $\bigl({}^{+1.8}_{-1.8}~~{}^{+2.3}_{-2.1}\bigr)$  & & 10.6  & & $5.7^{+2.6}_{-1.7}$  \\
    8j, $\geq$4{\cPqb}             & & $-0.2 $ & ${}^{+2.8}_{-3.0}$ & $\bigl({}^{+1.5}_{-1.5}~~{}^{+2.3}_{-2.6}\bigr)$  & & 5.5  & & $5.5^{+2.6}_{-1.6}$ \\
    $\geq$9j, $\geq$4{\cPqb} & & $-0.4 $ & ${}^{+2.1}_{-2.2}$ & $\bigl({}^{+1.4}_{-1.3}~~{}^{+1.6}_{-1.8}\bigr)$  & & 4.0  & & $4.3^{+1.9}_{-1.3}$ \\
    Combined          & & $0.9 $ & ${}^{+1.5}_{-1.5}$ & $\bigl({}^{+0.7}_{-0.7}~~{}^{+1.3}_{-1.3}\bigr)$  & & 3.8  & & $3.1^{+1.4}_{-0.9}$ \\
      \hline
    \end{tabular}
    \renewcommand{\arraystretch}{1.0}
\end{table}

\begin{figure}[hbtp]
  \centering
   \includegraphics[width=0.49\textwidth]{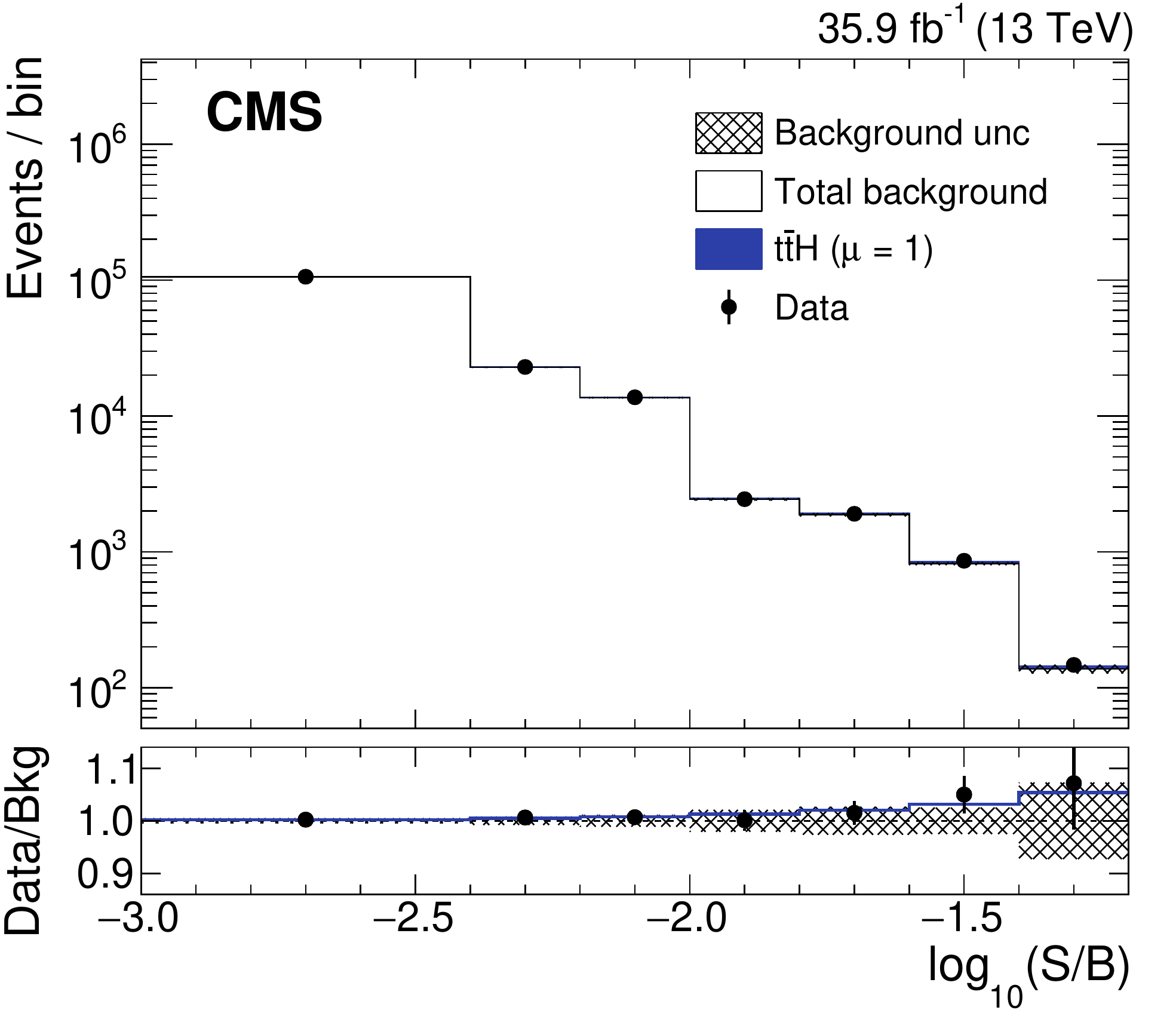}
   \caption{Distribution in $\log_{10}(\text{S}/\text{B})$, where S and B indicate the respective bin-by-bin yields of the signal and background expected in the MEM discriminant distributions, obtained from a combined fit with the constraint in the cross section of $\mu=1$.}
   \label{fig:log_sb}
\end{figure}

The signal is displayed in Fig.~\ref{fig:log_sb} in terms of the distribution in $\log_{10}(\text{S}/\text{B})$, where $\text{S}/\text{B}$ is the ratio of the signal to background yields in each bin of the six MEM discriminant histograms, obtained from a combined fit constrained to the SM cross section of $\mu=1$. Good agreement is found between the data and the sum of the SM signal and background over the whole range of this variable.

\clearpage 

\section{Summary}
\label{sec:summary}
A search for the associated production of a Higgs boson with a top quark pair is performed using proton-proton collision data collected by the CMS experiment at a centre-of-mass energy of 13\TeV, corresponding to an integrated luminosity of \lumifh.
Events are selected to be compatible with the \Hbb decay and the all-jet final state of the \ttbar pair, and are divided into six categories according to their reconstructed jet and {\cPqb} jet multiplicities.

The result of the search is presented in terms of the signal strength modifier $\mu$ for \ttH production, defined as the ratio of the measured \ttH production cross section to the one expected for a standard model Higgs boson with a mass of 125\GeV. From a combined fit of signal and background templates to the data in all event categories, observed and expected upper limits of $\mu < 3.8$ and $< 3.1$, respectively, are obtained at 95\% confidence levels.
These limits correspond to a best fit value of $\muhat = 0.9 \pm 0.7\stat \pm 1.3\syst = 0.9 \pm 1.5 \,(\text{tot})$, which is compatible with the standard model expectation.
\clearpage

\begin{acknowledgments}
We congratulate our colleagues in the CERN accelerator departments for the excellent performance of the LHC and thank the technical and administrative staffs at CERN and at other CMS institutes for their contributions to the success of the CMS effort. In addition, we gratefully acknowledge the computing centres and personnel of the Worldwide LHC Computing Grid for delivering so effectively the computing infrastructure essential to our analyses. Finally, we acknowledge the enduring support for the construction and operation of the LHC and the CMS detector provided by the following funding agencies: BMWFW and FWF (Austria); FNRS and FWO (Belgium); CNPq, CAPES, FAPERJ, and FAPESP (Brazil); MES (Bulgaria); CERN; CAS, MoST, and NSFC (China); COLCIENCIAS (Colombia); MSES and CSF (Croatia); RPF (Cyprus); SENESCYT (Ecuador); MoER, ERC IUT, and ERDF (Estonia); Academy of Finland, MEC, and HIP (Finland); CEA and CNRS/IN2P3 (France); BMBF, DFG, and HGF (Germany); GSRT (Greece); NKFIA (Hungary); DAE and DST (India); IPM (Iran); SFI (Ireland); INFN (Italy); MSIP and NRF (Republic of Korea); LAS (Lithuania); MOE and UM (Malaysia); BUAP, CINVESTAV, CONACYT, LNS, SEP, and UASLP-FAI (Mexico); MBIE (New Zealand); PAEC (Pakistan); MSHE and NSC (Poland); FCT (Portugal); JINR (Dubna); MON, RosAtom, RAS, RFBR and RAEP (Russia); MESTD (Serbia); SEIDI, CPAN, PCTI and FEDER (Spain); Swiss Funding Agencies (Switzerland); MST (Taipei); ThEPCenter, IPST, STAR, and NSTDA (Thailand); TUBITAK and TAEK (Turkey); NASU and SFFR (Ukraine); STFC (United Kingdom); DOE and NSF (USA).

\hyphenation{Rachada-pisek} Individuals have received support from the Marie-Curie programme and the European Research Council and Horizon 2020 Grant, contract No. 675440 (European Union); the Leventis Foundation; the A. P. Sloan Foundation; the Alexander von Humboldt Foundation; the Belgian Federal Science Policy Office; the Fonds pour la Formation \`a la Recherche dans l'Industrie et dans l'Agriculture (FRIA-Belgium); the Agentschap voor Innovatie door Wetenschap en Technologie (IWT-Belgium); the F.R.S.-FNRS and FWO (Belgium) under the ``Excellence of Science - EOS" - be.h project n. 30820817; the Ministry of Education, Youth and Sports (MEYS) of the Czech Republic; the Lend\"ulet ("Momentum") Programme and the J\'anos Bolyai Research Scholarship of the Hungarian Academy of Sciences, the New National Excellence Program \'UNKP, the NKFIA research grants 123842, 123959, 124845, 124850 and 125105 (Hungary); the Council of Science and Industrial Research, India; the HOMING PLUS programme of the Foundation for Polish Science, cofinanced from European Union, Regional Development Fund, the Mobility Plus programme of the Ministry of Science and Higher Education, the National Science Center (Poland), contracts Harmonia 2014/14/M/ST2/00428, Opus 2014/13/B/ST2/02543, 2014/15/B/ST2/03998, and 2015/19/B/ST2/02861, Sonata-bis 2012/07/E/ST2/01406; the National Priorities Research Program by Qatar National Research Fund; the Programa Estatal de Fomento de la Investigaci{\'o}n Cient{\'i}fica y T{\'e}cnica de Excelencia Mar\'{\i}a de Maeztu, grant MDM-2015-0509 and the Programa Severo Ochoa del Principado de Asturias; the Thalis and Aristeia programmes cofinanced by EU-ESF and the Greek NSRF; the Rachadapisek Sompot Fund for Postdoctoral Fellowship, Chulalongkorn University and the Chulalongkorn Academic into Its 2nd Century Project Advancement Project (Thailand); the Welch Foundation, contract C-1845; and the Weston Havens Foundation (USA).
\end{acknowledgments}

\bibliography{auto_generated} \cleardoublepage \appendix\section{The CMS Collaboration \label{app:collab}}\begin{sloppypar}\hyphenpenalty=5000\widowpenalty=500\clubpenalty=5000\vskip\cmsinstskip
\textbf{Yerevan Physics Institute,  Yerevan,  Armenia}\\*[0pt]
A.M.~Sirunyan,  A.~Tumasyan
\vskip\cmsinstskip
\textbf{Institut f\"{u}r Hochenergiephysik,  Wien,  Austria}\\*[0pt]
W.~Adam,  F.~Ambrogi,  E.~Asilar,  T.~Bergauer,  J.~Brandstetter,  E.~Brondolin,  M.~Dragicevic,  J.~Er\"{o},  A.~Escalante Del Valle,  M.~Flechl,  M.~Friedl,  R.~Fr\"{u}hwirth\cmsAuthorMark{1},  V.M.~Ghete,  J.~Grossmann,  J.~Hrubec,  M.~Jeitler\cmsAuthorMark{1},  A.~K\"{o}nig,  N.~Krammer,  I.~Kr\"{a}tschmer,  D.~Liko,  T.~Madlener,  I.~Mikulec,  E.~Pree,  N.~Rad,  H.~Rohringer,  J.~Schieck\cmsAuthorMark{1},  R.~Sch\"{o}fbeck,  M.~Spanring,  D.~Spitzbart,  A.~Taurok,  W.~Waltenberger,  J.~Wittmann,  C.-E.~Wulz\cmsAuthorMark{1},  M.~Zarucki
% ...
\vskip\cmsinstskip
% ...
1:~Also at Vienna University of Technology,  Vienna,  Austria\\
% ...
\end{sloppypar}
\end{document}